\documentclass[graybox]{svmult}

\usepackage{mathptmx}       % selects Times Roman as basic font
\usepackage{helvet}         % selects Helvetica as sans-serif font
\usepackage{courier}        % selects Courier as typewriter font
\usepackage{type1cm}        % activate if the above 3 fonts are
\usepackage{makeidx}         % allows index generation
\usepackage{graphicx}        % standard LaTeX graphics tool
\usepackage{multicol}        % used for the two-column index
\usepackage[bottom]{footmisc}% places footnotes at page bottom
\usepackage{hyperref}        %for hyperlinks
\usepackage{soul}            % for high-lighting of text
\usepackage{xcolor}             % colored text 
\hypersetup{colorlinks=true,urlcolor=blue}
\usepackage[square,numbers]{natbib}
\DeclareRobustCommand{\ion}[2]{\textup{#1\,\textsc{\lowercase{#2}}}}
\newcommand*\element[1][]{%
  \def\aa@element@tr{#1}%
  \aa@element
}

\def\ch{{\it Chandra}}
\def\xmm{XMM-{\it Newton}}
\makeindex             % used for the subject index
\begin{document}
%\tableofcontents{}
\title*{Interstellar absorption and dust scattering}
\author{E. Costantini \thanks{corresponding author} and L. Corrales}
\institute{First Author \at SRON Netherlands Institute for Space Research, Niels Bohrweg 4, 2333CA Leiden, The Netherlands, \email{e.costantini@sron.nl}
\and Second Author \at University of Michigan, Dept. of Astronomy
1085 S University Ave
Ann Arbor, MI 48109, USA \email{liac@unimich.edu}}
\maketitle
\abstract{The study of the interstellar medium (ISM) in the X-rays has entered a golden age with the advent of the X-ray observatories \xmm\ and \ch. High-energy resolution allowed to study dust spectroscopic features with unprecedented detail. At the same time, the X-ray imaging capabilities offered a new perspective of dust scattering halos. Both spectroscopy and imaging rely on a simple geometry, where a distant X-ray source, usually a bright X-ray binary system, lies behind a multi-layered ISM. X-ray binaries can be found in different regions in the Galaxy, providing the unique chance to study the ISM in distinct environments. In the following we will describe how X-rays can be used as a tool to study gas and dust along the line of sight, revealing elemental abundances and depletion. The study of interstellar dust spectroscopic and imaging features can be used to extract the chemical and physical properties of the intervening dust, as well as its distribution along the line of sight.}

\vspace{0.3cm}

\section{Introduction}
The interstellar medium is an environment with densities spanning many orders of magnitude ($\sim10^{-4}-10^{6}$\,cm$^{-3}$, \citep[][]{mckee77,ferriere01}), filling the space not occupied by stars in our Galaxy. This medium has been recognized as a substance in-between stars already in the XVII century \citep[][]{bacon1626}. Its nature and properties caused a growing interest over the centuries which brought prominent astronomers of their time to try to characterize this medium \citep[e.g.,][]{barnard,trumpler}. Today, the interstellar medium, which constitutes about 10\% of the Galaxy's visible mass, is one of the most extensively studied fields of astronomy.
Similar to the stars, the ISM composition is by far dominated by hydrogen. Depending on its ionization, the medium can be roughly divided in three phases. Neutral hydrogen, \ion{H}{I}, traces about 60\% of the ISM \citep[e.g.,][]{draine11,mckee77}. The temperatures encompassed by this diffuse medium are in the range 100--5000\,K. A colder phase ($T$=10--50\,K) accounts for dense media and molecular clouds. Its importance in terms of volume in the ISM is more modest, as about 17\% of hydrogen is in the H$_2$ form that characterize these media. Finally, environments with temperatures $>10^{4}$\,K are dominated by \ion{H}{II}. The gas in these regions are either photoionized by nearby stars or heated by shock events in the medium.

The majority of interstellar material production is due to stellar activity \citep{clayton78a}. Dense environments, like molecular clouds, may collapse due to their own gravity to eventually form a star. After the star has evolved along the main sequence of the Hertzprung-Russell diagram and turns into a giant star, the outer layers, containing gas from the recent dredge-up events, as well as gas from the pristine material (e.g. C, O, Si and Mg), are ejected and returned to the surrounding medium \citep[e.g.,][]{whittet03}. As the temperature decreases with distance from the star \citep[][]{ss96}, gas may condense into solid grains (Sect.~\ref{par:star}).

If a star has a sufficiently high initial mass ($>8\,M_\odot$), the giant-star phase will end into a core-collapse supernova (CCSN) explosion. In these conditions, iron can be formed either by the short-lived Si-burning or by decay of the $^{56}$Ni. Actual dust formation is believed to take place in ejecta of CCSN \citep[][]{dwek16,slavin20}. The analysis of young remnants pointed out that the aftermath of the explosion can produce up to $\sim0.4\,M_\odot$ of dust \citep[e.g., SN1987a,][]{matsuura11}. Similar values, summing the silicates and carbon contribution, 
were recently reported for the supernova remnant Cas\,A \citep[$0.4-0.6\,M_\odot$,][]{delooze17}, while $\sim0.22\,M_\odot$ was reported for the Crab nebula \citep[][]{gomez12}. 
Type\,Ia SN should be also efficient producers of iron \citep[][]{nozawa11}, formed in the innermost part of the explosion region of a white dwarf which exceeded the Chandrasekhar limit. However, iron dust from SN type\,Ia has been calculated to be readily destroyed by the SN reverse shock, therefore returning into gas phase \citep[e.g.,][]{gomez12b}.  In summary, the late stage of the star life is fundamental in the dust and gas cycle in the ISM, providing an efficient way to newly produce interstellar material.

In the following we describe the general properties of the cold ISM, in particular focusing on those aspects interesting for X-ray investigation. This will be necessarily concise. Then we illustrate the phenomenology of the ISM as seen by the current X-ray instrumentation. This is followed by a more quantitative description of the physical processes involved. A part of this chapter describes how new models for interstellar dust modeling are developed: from the laboratory measurements to the implementation into fitting routines. For clarity we treat separately the ISM extinction as seen from high-resolution X-ray spectroscopy and dust scattering, studied with CCD imaging. At the end, the state of art of our current understanding of the ISM from the X-ray point of view can be found, followed by an outlook on future missions.  

\section{The cold ISM}
The spiral arms pattern in our Galaxy may be recovered from the study of the distribution and properties of the neutral hydrogen emission \citep[][]{levine06}, the \ion{H}{II} distinct regions, and molecular clouds, highlighted by H$_2$ \citep[][]{levine06,kalberla07}. The scale height  of the disk occupied by diffuse emission changes considerably with distance from the center, going from 0.15\,kpc in the central regions, up to $\sim2.2$\,kpc at the outer end of the Galactic profile traced by neutral hydrogen \citep[$\sim$35\,kpc,][]{kalberla09}. The column density of the neutral hydrogen changes widely in the Galaxy and typically spans more than two orders of magnitudes, around the range $\sim\,10^{20-22}$\,cm$^{-2}$. Extreme values at both ends of this interval can be found. As soon as the \ion{H}{I} column density exceeds $\sim4\times10^{20}$\,cm$^{-2}$, the diffuse medium may superimpose with the colder, molecular medium, mainly traced by the CO molecule\citep[][]{combes91,dame01}. Contrary to the diffuse medium, the distribution of molecular material is clumpy. Often warm clumps lie at preferential longitudes, overlapping with the \ion{H}{II} regions population, following the spiral structure \citep[e.g.,][]{solomon85}. The cold phases of hydrogen witness therefore both the star formation and the release of the interstellar material in the diffuse medium. Out of this reservoir come the material for new stars and planets formation.

\subsection{Interstellar dust}\label{par:star}

Intertwined with the cold gas is the dust phase, which constitutes about 1\% of the total ISM \citep[][]{whittet03,boulanger00}. The presence of solid dust particles has been soon recognized, as the abundance of certain elements, measured from UV absorption lines, appeared significantly sub-solar \citep[e.g.,][]{ss96}. Abundant elements, like carbon, oxygen and iron appeared significantly depleted from the gas phase, indicating that they should be present in another form, for example in dust grains. 
The depletion of a given element $x$ can be denoted as  $D(x)$, as \citep[e.g.,][]{whittet03}:
\begin{equation}
D(x)=log\left[\frac{N_x}{N_H}\right]-log\left[\frac{N_X}{N_H}\right]_{\odot}.
\end{equation}
This can be simply transformed into the fractional depletion $\delta(x)$:
\begin{equation}
\delta(x)=1-10^{D(x)}.
\end{equation}
Where $\delta(x)=1$ indicates that the element is totally included in dust, while if the element is only in gas form, $\delta(x)=0$.  
The depletion value is a function of temperature \citep[][]{ss96} and, as a consequence, of the environment where the element resides. In general, the denser and colder the environment, the more depleted an element is \citep[][]{jenkins09}.

A large amount of observational multi-wavelength (from radio to far-ultra-violet) evidence on dust has been collected in the last decades, allowing the determination of the general chemical composition of interstellar dust, its size, shape and physical characteristics.\\ 
The broad band spectral energy distribution \citep[e.g.][]{compiegne11}, the extinction curve (Sect.~\ref{par:extcurve}), as well as detailed infrared (IR) and far-infrared spectroscopy \citep[e.g.,][and references therein]{molster10} determined that carbon and silicates should dominate the chemical composition of dust. Carbon, produced in the aftermath of wind ejection from carbon-rich red-giant stars, has historically been assumed to be in the form of graphite. The prominent and ubiquitous extinction feature at 2175\,\AA\ (Fig.~\ref{f:ext}) can be interpreted as coming from small graphite grains, whose excitation energy would be consistent with the position of the observed feature \citep{stecher_donn}. However, the line-of-sight broadening variations have been proposed to be due to the same type of excitation, but from the polycyclic-aromatic-hydrocarbons (PAH) molecules \citep{draine03a}.    
Graphite, in analogy with silicates, should also face a process of amorphisation in the ISM. Therefore amorphous carbon or hydrogenated carbon (HAC), have been proposed as a possible candidates for the C reservoir \citep[e.g][]{compiegne11, duley89}.\\

The inclusion of Mg, Si, Fe and O into silicates has been proven through IR spectroscopy. The 9.7\,$\mu$m and 18\,$\mu$m absorption features, seen in the environment of oxygen rich stars, have been indeed interpreted as the stretching and bonding modes of Si--O and O--Si--O, respectively \citep[e.g. ][]{draine03a,molster10}.\\
The amount of iron and magnesium inclusion in the silicate depend on the initial conditions in the giant-star winds where they were formed \citep[][for a review]{gail10}. As the gas flows farther from the star and temperatures decrease with distance, the first dust particles can be formed by condensation, starting from Ti, at $\sim1500$\,K. According to calculations assuming chemical equilibrium \citep[Fig.~9 in][]{gail10}, in a circumstellar envelope Mg and Si should bind into olivine first ([Mg,Fe]$_2$SiO$_4$) at around 1100\,K. Then, in rapid succession, pyroxene ([Mg,Fe]$_{0.5}$SiO$_3$ at 1000\,K) and metallic iron (at around 900\,K) should form.  This condensation sequence is indeed noticed in high mass-flow rate stars \citep{molster10}. In this case, the crystalline Mg-rich end of the olivine and pyroxene (Mg$_2$SiO$_4$ and MgSiO$_3$, respectively) are observed. The fraction of crystalline dust in those environment is relatively small \citep[10--15\%,][]{molster10}.
However, dust grains are often formed in a fast-cooling and evolving outflows, where the equilibrium conditions just described do not apply. 
In non-equilibrium, when the temperature becomes lower than $900$\,K, the dust forms preferentially amorphous aggregates, observed to be iron-rich \citep[e.g. in protostars,][]{demyk99}.

Both Mg and Si in solid form are almost completely included in silicates \citep{whittet03}. However the high depletion of iron ($>90\%$) cannot be explained only in terms of silicate inclusion \citep[][]{jenkins09,dwek16,poteet15,svitlana18}. 
Iron can also exist in other stable forms, as metallic iron, iron oxides, and iron sulfides \citep[e.g.][]{svitlana18}. The presence of metallic iron cannot be directly observed at long wavelengths, as iron should not display any vibrational mode \citep[][]{molster10}, making it difficult to directly test the presence of this iron form. It has been hypothesized that iron in FeS (with small inclusion of Ni) could exist as inclusion into larger silicates grains, mainly formed by amorphous enstatite. These aggregates are called Glasses with Embedded Metals and Sulfides \citep[GEMS,][]{bradley94}. They are commonly observed in comets and they are a constituent of the interplanetary dust particle reservoir.   
Therefore, the majority (90--99\%) of GEMS particles should not have an ISM origin, but are believed to be formed in the solar nebula itself \citep[][]{keller11}. However, GEMS-like grains may constitute a fraction of the amorphous silicate grains in the ISM. Indeed, GEMS with anomalous oxygen isotopic composition may have been processed in the ISM \citep[][]{messenger03}. Recent detailed studies on cometary GEMS, point out that they must have undergone more than one stage of processing, in a cold environment. The detected organic carbon in those GEMS would not indeed survive in the hot environment ($>1300$\,K) of the solar nebula \citep[][]{ishii18}. The Cassini mission, in orbit around Saturn, allowed the detection of dust grains consistent to be reprocessed multiple times in the ISM. The composition has been found to be dominated by magnesium-rich silicates with iron inclusion \citep[][]{altobelli16}.

Lower abundance metals are often highly depleted \citep[e.g.][]{jenkins09}. For example Al, Ca, and Ti, among the first elements to condense in the stellar envelope (at $T=1400–1600$\,K, \citep[][]{ss96,field74}) are believed to form the first and innermost core of more complex silicate grains \citep[][]{clayton78}. This inclusion would provide a natural protection and would explain why the high depletion of these elements is almost insensitive to the environment temperature. Calcium carbonates have also been reported in spectra of envelopes of asymptotic-giant-branch stars \citep[e.g.,][]{kemper02}. Another low-abundance element, nickel, has a depletion pattern similar to the one of iron and a similar condensation temperature (1336 and 1354 K, for Fe and Ni, respectively) suggesting a common dust inclusion history. In equilibrium conditions, they should indeed condense into nickel-iron in the stellar envelope \citep[][]{gail10}.

The temperature of the environments where dust resides has a profound impact on the grain internal structure. In the inner part of the stellar outflow, the newly formed dust is crystalline, thanks to the high temperatures \citep[][]{gail10}. However, any dust formed at temperatures $<900$\,K, will be the consequence of a disordered aggregation, forming glass-like structures. Annealing, i.e. the process of reordering the lattice internal structure, can only happen if the dust is again exposed to a high temperature for a sufficiently long amount of time. This is the case for comets, where crystalline silicates have been detected \citep[][for a review]{hanner04}. Crystalline dust has been also associated with protoplanetary disks \citep[e.g.][]{watson09}. However, in the diffuse ISM, because of the low temperatures, dust is observed to be mostly amorphous. From {\it Spitzer} observations along diffuse sightlines, the amount of crystalline dust has been determined to be tiny: less than 2\% according to \citet[][]{kemper04} or less than 5\% \citep[][]{li07}.

\subsection{The extinction curve}\label{par:extcurve}
\begin{figure}[t]
    %\centering
    \includegraphics[angle=0,width=1.0\textwidth]{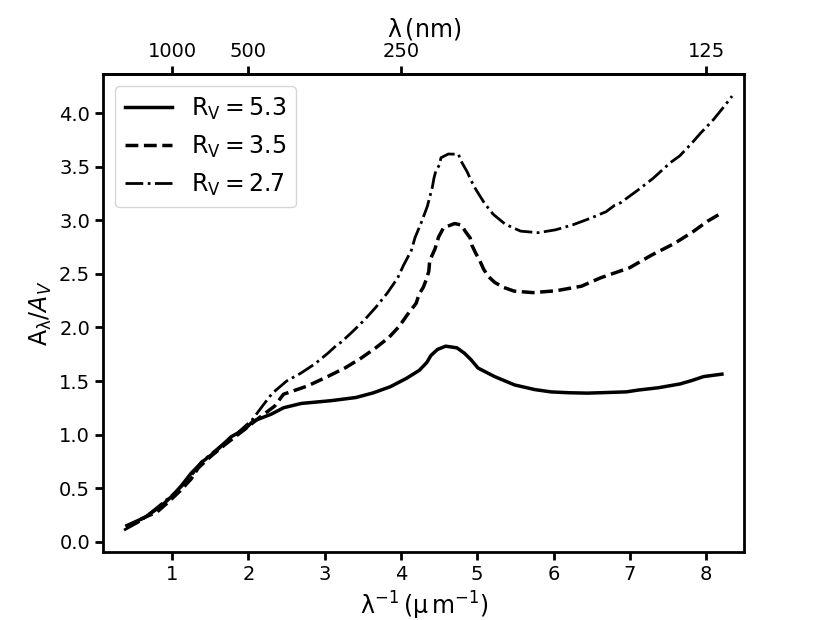}
    \caption{Examples of observed extinction curves at different values of $R_V$. The data are related to Herschel\,36 (solid line), HD\,48099 (dashed line) and BD+56\,524 (dashed-dotted line). Adapted from \citet{cardelli89}.}
    \label{f:ext}
\end{figure}

The information given by the extinction curve, i.e. how the background star light is attenuated by intervening material as a function of wavelength, help determine, together with other properties, the dust size distribution \citep[e.g.][]{tielens13}. The extinction curve is measured in terms of extinction ratio $A_{\lambda}/A_V$ (Fig.~\ref{f:ext}), where the extinction, measured in magnitudes, indicates how much of the star light is removed when observing a background star. Thus, measuring the difference in extinction between the observed star and the theoretical magnitude a star should have, given a spectral type and luminosity class, provides the color excess:\\
\begin{equation}
E(B-V)=(B-V)_{observed}-(B-V)_{intrinsic},    
\end{equation}
where $B$ and $V$ are the magnitudes in the blue (centered at 442\,nm) and visual (centered at 540\,nm) photometric bands, respectively. 
The value of the 
total to selective extinction ratio $R_V=A_V/E(B-V)$ is used to characterize the environment (Fig.~\ref{f:ext}). The typical value for diffuse sight-lines is $R_V=3.1$, while dense environments (e.g. star formation regions) show a larger value of $R_V$ \citep{tielens13}, resulting in a flatter extinction curve, where the value of $A_V/A_\lambda$ is less sensitive to the wavelength.

The visual extinction $A_V$ has been found to well correlate to the amount of neutral hydrogen in the ISM, confirming that cold gas and dust are in general well mixed. The relation between the amount of hydrogen, traced by the H\,Ly$\alpha$ and H$_2$ absorption, and $A_V$ has been found to be  $N_{\rm H} ({\rm cm}^{-2})\sim 1.87 \times 10^{21} A_V$\,(mag) \citep{bohlin78}. This relation has been confirmed by subsequent comparisons between extinction and both the H\,Ly$\alpha$ \citep[e.g.][]{whittet81} and the \ion{H}{I}\,21\,cm line \citep[e.g.][]{liszt14}, albeit with slightly different slopes 
\citep[][for a comprehensive list of references]{zhu17}.

Ever since the first X-ray missions were online \citep[e.g.][]{gorenstein75}, the same trend has been confirmed using the hydrogen column density as measured in the X-rays. Measurements of the hydrogen column density towards X-ray binaries brought the relation:    
$N_{\rm H} ({\rm cm}^{-2})\sim 1.79 \times 10^{21} A_V$\,(mag) \citep[using the ROSAT satellite,][]{Predehl1995}. Further studies using X-ray binaries \citep{valencic15,zhu17}, supernova remnants, and gamma ray bursts \citep{watson}, observed with different X-ray instruments, including {\it Chandra} and XMM-{\it Newton}, reported a steeper slope: $N_{\rm H} ({\rm cm}^{-2})\sim 2.08-2.87 \times 10^{21} A_V$\,(mag) \citep[][]{guver09,foight16,zhu17}. 
These relations are affected by some scatter, but all confirm the tight connection between gas and dust in the diffuse ISM both along short \citep[1-2\,kpc,][]{bohlin78} and long \citep[up to $\sim$14\,kpc,][]{Predehl1995} sightlines. 

\subsubsection{The dust size distribution}\label{s:size}
The extinction curve (Fig.~\ref{f:ext}) cannot be explained by a single-size grain population, but rather a distribution of sizes. One of the simplest formulation of the dust size distribution prescribes that the number of grains ($n$) follows a powerlaw distribution as a function of the grain radius \citep[MRN distribution, ][]{Mathis1977}:

\begin{equation}
n(a)\,da = A\, n_H\, a^{- 3.5}da.
\end{equation}

This relation is valid for $a=0.005-0.25$\,$\mu$m. The term $n_H$ is the number density and $A$ is a normalization factor, which assumes slightly different values for silicates and carbonaceous grains. In particular, $A_S = 1.73\times10^{-25}$ and $A_C = 1.62\times10^{-25}$\,cm$^{2.5}$ \citep{Mathis1977,Mauche1986}. The MRN model is used in a number of applications, as it is a simple analytical function. However refinements and extensions of this model have been investigated. \citet[][]{zubko04}, starting from the observational constraints of the extinction curve, the diffuse IR emission and the elemental abundance, describe the dust size distribution for different chemical compositions. More than one solution in that study was found to fit well the data, implying that additional limits from other observables are necessary to constrain the dust size distribution models.\\
The small- and large grain ends of the dust size distribution have been further studied. The smaller grains inclusion has been presented in e.g., \citep[][]{wd01}. For both silicates and carbonaceous grains, the lower limit on the grain radius was extended down to 3.5\,\AA, therefore including the contribution of PAH \citep[][]{tielens08} for carbon, missing from earlier formulations.

Studies of starlight, polarized as it passes the ISM through multilayered dust clouds, help constraining instead the large grain population. Small grains do not contribute significantly to the polarized light, therefore they should be either spherical or not aligned with the magnetic field \citep[e.g.][and references therein]{df09}. The grains which polarize light are most probably prolate in shape (i.e. an elongated rugby ball), whose smaller axis is mostly aligned with the magnetic field, in a configuration that has been dubbed as ``picket-fenced" \citep{db74}. The mean radius of the polarizing grains has been found to be near 0.1\,$\mu$m, reaching up to 0.2--0.3\,$\mu$m, depending on the line of sight \citep[][]{siebermorgen18}.

One mechanism to form larger grains is by aggregating smaller grains \citep[][ and references therein]{hirashita14}. An indirect evidence of coagulation come from the extinction curve, becoming flatter as the parameter $R_V$ increases (Fig.~\ref{f:ext}). This coagulation is expected to be disordered, resulting in aggregates with voids, called ``fluffy" grains. 

\section{Attenuation of X-rays by the Interstellar Medium}\label{par:attenuation}

Evidence of the ISM influencing the observed X-rays from astrophysical objects has been reported ever since the time of the first X-ray space missions. Both HEAO-1  \citep[][]{charles79} and {\it Einstein} \citep[][]{sc86} attempted a measurement of both the hydrogen and oxygen column density of the ISM towards the line of sight of the Crab X-ray emitter. Fast forward to this millennium, the X-ray observatories XMM-{\it Newton} \citep{jansen01} and {\it Chandra} \citep{weisskopf00} with their multiple instruments on board, allowed us a much deeper understanding of extinction in the X-ray band.

The intrinsic spectral shape of the X-ray background source appears attenuated according to: $I(E)=I_0(E) {\rm exp}(-N_{\rm H}\sigma(E))$ where $I_0$ is the original spectral shape, $N_{\rm H}$ is the column density and $\sigma(E)$ it the energy dependent extinction cross section. The total gas cross section at a given X-ray energy can be given by the sum of the individual element contribution, weighted by its cosmic abundance, ionization state and depletion \citep[][]{ride77,wilms00}:
\begin{equation}
\sigma_{tot}=\sum_{Z,i} A_Z \times \alpha_{Z,i} \times (1-\beta_{Z,i}) \times \sigma_{Z,i}.
\end{equation}
For each element $Z$, $A$ is the abundance, while $\alpha$ is the fraction of ions at ionization $i$. The term $(1-\beta_{Z,i})$ is the depletion of the ion $i$, defined as the ratio of the gas over the total ISM material (gas and dust). Finally,  $\sigma_{Z,i}$ is the individual ionic cross section. The atomic cross section as a function of energy decreases in first approximation as $E^{-3}$ \citep[e.g.,][]{band90}

\begin{figure}
    \centering
    \includegraphics[width=1.2\textwidth]{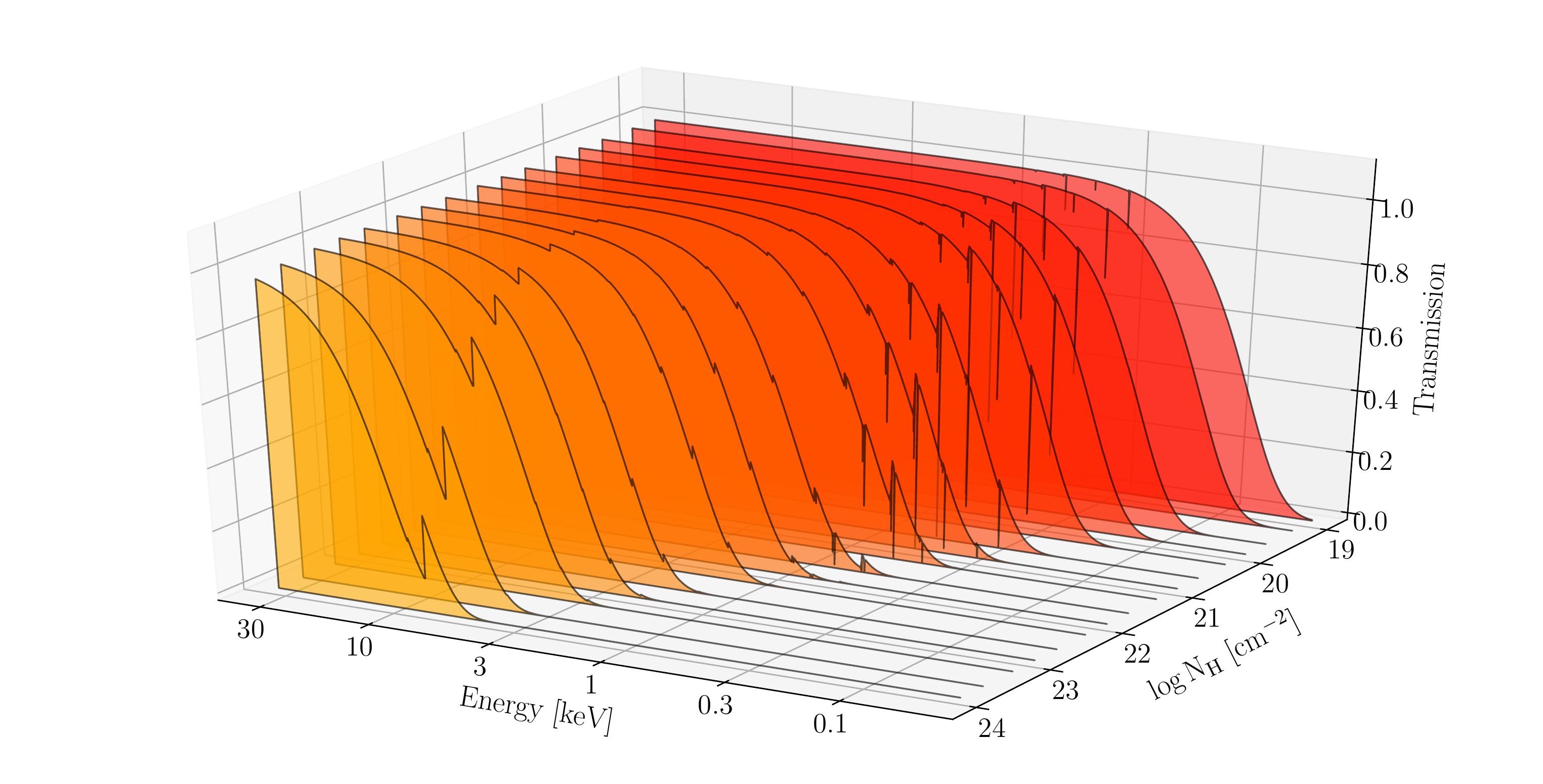}
    \caption{Transmitted X-ray spectrum as a function of energy and column density \citep{rogantini_thesis}.}
    \label{fig:nh}
\end{figure}

The elements represented by significant spectroscopic features in the X-ray band are: C, N, O, Ne, Mg, Si and Fe. Absorption structures that even with lower resolution spectrometers could be disentangled are the bound-free transitions from the K-shell (and sometimes also from the L-shell). They can be simply described, at $E>E_0$, by an `edge' function of the form $exp(\tau_0(E/E_0)^{-3})$, where $E_0$ is the threshold X-ray energy at which the electrons can be expelled from the shell and $\tau_0$ is the absorption depth at energy $E_0$. Bound-bound transitions of neutral or mildly ionized ions in the ISM can also be present as well as absorption by highly ionized gas, described elsewhere in this volume. They clearly involve less energy than the absorption edge of the same ion, therefore they appear in the spectrum generally at lower energies (or larger wavelength $\lambda=12.3985/E$, if $\lambda$ is expressed in \AA\ and $E$ in keV). In the ISM however, gas and dust always coexist in cold environments. The effect of dust in an X-ray spectrum can be noticed in three ways: $(i)$ depletion of the gas phase into dust, resulting in an apparent underabundance of a given element, namely the ones locked in dust (Sect.~\ref{par:star}). $(ii)$ The effect of X-rays interacting with solid particles, rather than gas, resulting in resonance effects (Sect.~\ref{par:xafs}). $(iii)$ The effect of scattering of the X-rays by the dust solid particles, which changes the overall cross section slope as a function of energy \citep[][and Sect.~\ref{par:scattcorrection}]{Draine2003}. In Fig.~\ref{fig:nh} the effect of the cold ISM column density $N_{\rm H}$ on a transmitted X-ray spectrum is shown. The smooth cut-off at lower energies is due to long tail of the hydrogen deep absorption edge at 13.6\,eV. This curvature can be accurately measured, especially by broad band X-ray spectrometers, and the hydrogen column density determined \citep[e.g.][]{kaastra08}. Albeit at a lower level, also He, \ion{H}{II} and H$_2$ contribute to the low-energy cut-off.

\subsection{Dust Scattering from the ISM}
\label{sec:DustScattering}

X-ray scattering has been used to study crystalline structures in different materials since the beginning of the XX century. The first time that this phenomenon was brought to attention regarding X-ray propagation through the interstellar medium was by \citet{Overbeck1965}, who noted that the apparent size of X-ray sources should increase due to the presence of interstellar dust. The source should then be surrounded by a round, diffuse halo. \citet{Hayakawa1970} showed that this phenomenon could be used to measure the distance, size, and composition of interstellar dust grains. The first observation of a so called ''dust scattering halo" was not achieved until over a decade later, around the bright high mass X-ray binary GX~339-4, imaged with the \textit{Einstein} Observatory \citep{Rolf1983}. The first survey of dust scattering halos was enabled by the launch of the ROSAT satellite, which performed an all sky survey in the X-rays, revealing scattering halos around 25 bright point sources and providing benchmarks for scaling relations between optical and X-ray extinction properties of the ISM \citep{Predehl1995}. 
Since then, dozens of dust scattering halos have been studied in detail, placing constraints on the dust grain size distributions in the ISM as well as the location of dust clouds along the sight line to background Galactic X-ray binaries \citep[see][ for canonical examples]{Smith2002, DraineTan2003, Costantini2005, Smith2008}.

\begin{figure}
\begin{center}
    \includegraphics[width=0.75\textwidth]{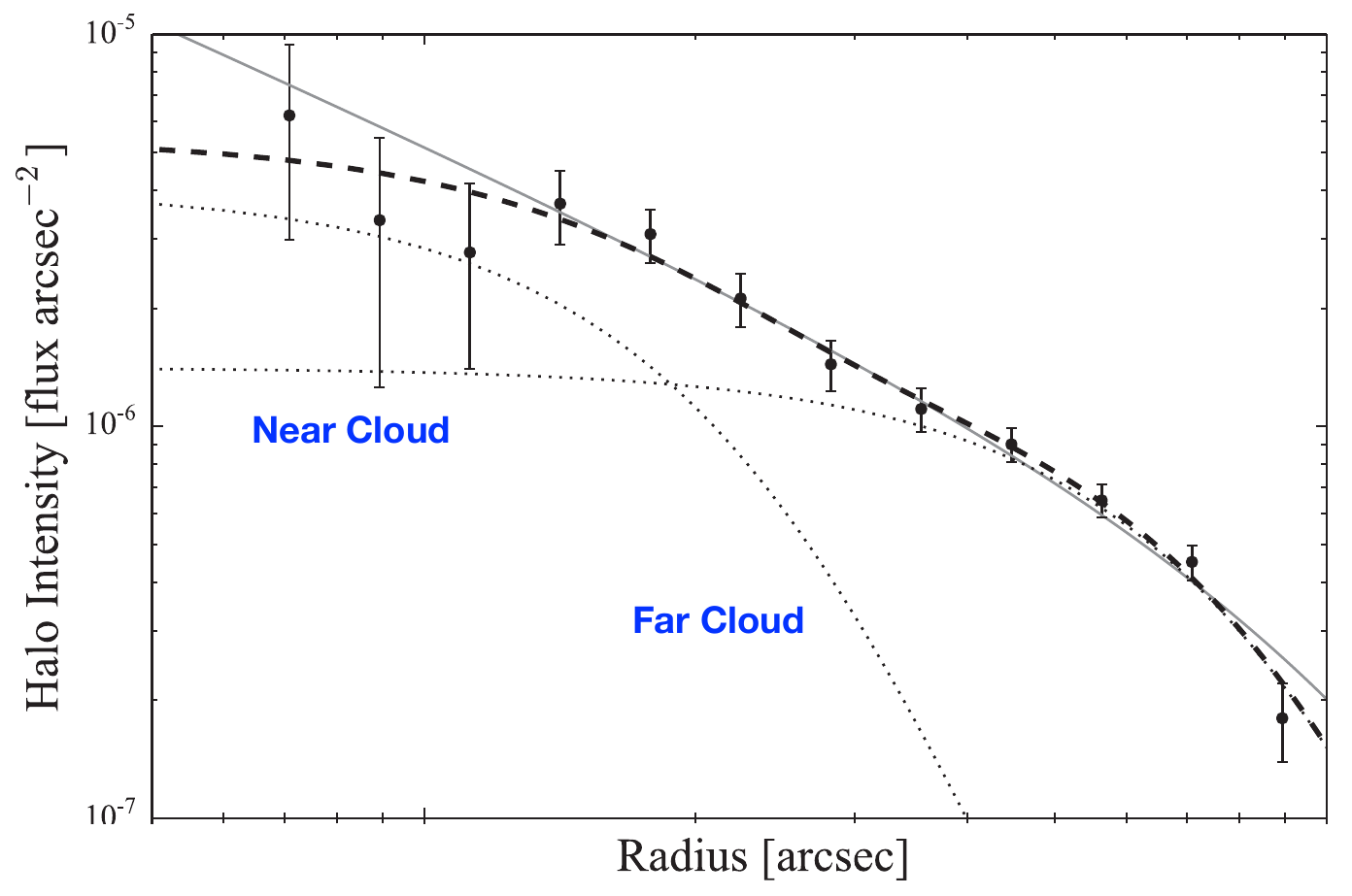}
    \caption{
    The 4-6 keV surface brightness profile of the dust scattering halo around Cygnus X-3. The radial profile can be fit with a model of dust that is distributed uniformly along the line of sight (grey solid line) or by assuming that the dust is concentrated in two dust clouds (dotted lines) at intermediate distance between the observer and the X-ray source (black dashed line). The data were gathered by the \ch\ satellite. Modified from \citep{CorralesPaerels2015}. 
    }
\label{fig:ScatteringExample}
\end{center}
\end{figure}
Dust scattering halos are typically on the order of $10$~arcmin in angular extent, arise from dust approximately  located at intermediate distances between the bright X-ray source and the observer, and probe ISM regions with physical sizes on the order of $1-30$~pc along the line of sight. Figure~\ref{fig:ScatteringExample} shows an example of the surface brightness profile of the scattering halo that originated from the high-mass X-ray binary Cygnus X-3.

The theory of scattering by small particles is covered in detail by a several text books on the subject \citep{vandeHulst, BohrenHuffman}. In the case of X-rays, interstellar dust is relatively transparent to the incident radiation $(|m-1| << 1)$, where  $m$ denotes the complex index of refraction. 
Because the grains are typically much larger than the wavelength of incident radiation, there is a minimal 
phase shift when the light enters the particle $(2 \pi a \lambda^{-1} |m-1| << 1)$, where $a$ is the radius of the dust grain. These are the conditions required to apply the Rayleigh-Gans approximation, yielding:
$$ \sigma_{\rm sca} \cong 2 \pi a^2 \left( \frac{2 \pi a}{\lambda} \right)^2 |m-1|^2 . $$
By nature of the approximation, the differential cross-section can be calculated by assuming that the electromagnetic wave inside the dust grain is the same as that incident upon it and integrating the scattered wave-fronts from every part of the grain. For spherical dust grains, the differential scattering cross-section can be approximated with a Gaussian function \citep{Mauche1986}, 
$$ \frac{d\sigma_{\rm sca}}{d\Omega} \approx \frac{4 a^2}{9} \left( \frac{2 \pi a}{\lambda} \right)^4 |m-1|^2 \exp\left( -\frac{\theta_{\rm sca}^2}{2 \bar{\sigma}}\right). $$
Here $\Omega$ is the solid angle and $\theta_{\rm sca}$ is the scattering angle. 
Under the assumption of small scattering angles, $\bar{\sigma}$ can be approximated as:\\ 
$$ \bar{\sigma} = 10.4 \left( \frac{E}{{\rm keV}} \right)^{-1} \left( \frac{a}{0.1\mu{\rm m}} \right)^{-1}~{\rm arcmin}. $$ 
This provides a sense of the halo angular extend for a single grain size \citep{Mauche1986}.\\ 
An approximation for the dielectric response can be made by treating the dust grain as a collection of free electrons, the ``Drude approximation'' \citep{Smith1998}:
$$ |m-1| \approx \frac{n_e r_e \lambda^2}{2\pi} $$
where $n_e$ is the average density of electrons in the grain and $r_e$ is the classical radius of an electron. Applying the Drude approximation yields the canonical X-ray scattering cross-section:
$$ \sigma_{\rm sca} \approx 6.168 \times 10^{-11}~{\rm cm}^{2}\  
\left( \frac{\rho}{3~{\rm g\ cm}^{-3}} \right)^2 
\left( \frac{a}{0.1~\mu{\rm m}} \right)^4\ 
\left( \frac{E}{{\rm keV}} \right)^{-2} $$
where $\rho$ is the material density of the dust grain, $E$ is the energy of the incident light, and it is assumed that the material contains roughly equal numbers of protons and neutrons.\footnote{In other words, this assumes $n_e \approx \rho / 2 m_p$} The energy dependence of the cross-section demonstrates that scattering is more important for soft X-rays. The strong dependence on particle radius $(a^4)$ makes it so that the dust scattering phenomenon is particularly powerful for constraining the large end of the cosmic dust grain size distribution \citep{Witt2001, CorralesPaerels2015, Valencic2019WP}. An important caveat to this point is that the Rayleigh-Gans approximation breaks down as grains get larger ($a \geq 1~\mu{\rm m}$) or as we look towards very soft X-rays ($E \leq 0.3$~keV). \citet{Smith1998} demonstrate that Rayleigh-Gans scattering can be applied as long as the energy of the incident X-ray photon, in keV, is significantly larger than the grain radius, in microns. When the Rayleigh-Gans approximation no longer applies, one can employ the more general anamolous diffraction theory or Mie scattering theory (Section~\ref{par:mie}). Doing so demonstrates that the scattering cross-section for X-rays is roughly flat for $E({\rm keV}) \leq a (\mu{\rm m})$, and no longer follows the canonical $E^{-2}$ dependence \citep{Smith1998, CorralesPaerels2015}.

\begin{figure}
\begin{center}
\includegraphics[width=\textwidth]{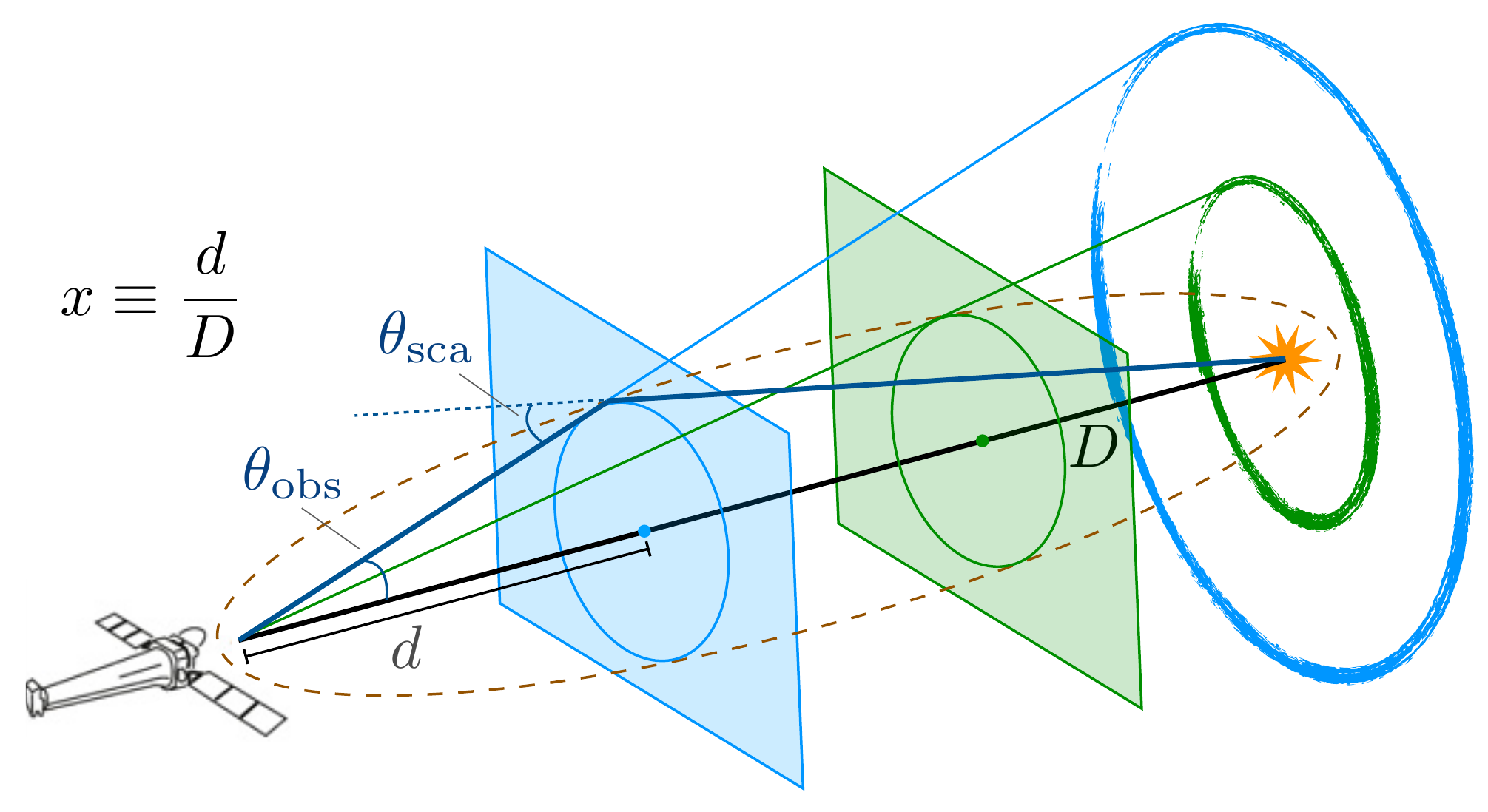}
\caption{Illustration of the geometric principles used to compute the dust scattering halo intensity under the assumptions of single scattering. The positions of two dust clouds are represented by the blue and green planes along the line of sight between a telescope and an X-ray source, separated by a total distance $D$. The apparent angular distance between the point source and dust-scattered light is represented by $\theta_{\rm obs}$. In order for scattered light to reach the observer, it must fall onto the angle $\theta_{\rm sca}$, which equals $\theta_{\rm obs}/(1-x)$ under the small angle approximation. In the case that the X-ray source undergoes a bright outburst, the observer will see scattering from dust that lies along equal path lengths, represented by the ellipsoid in the illustration. Where the ellipsoid intersects the dust clouds, a ring pattern is observed, growing in angular size with time. [Adapted from illustrations by S.~Heinz]}
\label{fig:HaloGeometry}
\end{center}
\end{figure}

To model the surface brightness profile of the resulting scattering halo image, one must integrate the differential scattering cross-section over the interstellar dust grain size distribution while accounting for geometrical effects of the sight line. Figure~\ref{fig:HaloGeometry} demonstrates the layout of the problem and defines fundamental parameters such as the dust fractional distance, $x \equiv d/D$, where $d$ is the distance to the dust particle and $D$ is the distance to the background X-ray source. Employing small angle approximations, the integral can be written as:
\begin{equation}\label{eq:sca}
\frac{dI}{d\Omega} (E, \theta_{\rm obs}) = \int_0^1 \int_{a_{\rm min}}^{a_{\rm max}} F_a(E)\ \frac{d\sigma_{\rm sca}}{d\Omega}(E, a, \theta_{\rm sca})\ n_d(a)\ \xi(x)\ da \ dx. 
\end{equation}
Here it is also assumed that the sight-line is optically thin, therefore only one scatter occurs before reaching the observer. 
In Eq.~\ref{eq:sca}, $F_a$ is the absorbed (apparent) flux of the background X-ray source after the effect of ISM absorption, $n_d$ is the grain size distribution (assumed to be the same everywhere along the sight line), $\xi(x)$ represents the density distribution of dust along the sight line, and the differential scattering cross-section must be evaluated for the appropriate scattering angle, $\theta_{\rm sca} = \theta_{\rm obs}/(1-x)$. 
Various approximations and semi-analytical solutions for the scattering halo intensity profile can be found in the literature \citep{Mauche1986, Draine2003, CorralesPaerels2015}. 
The scattering halo integral above also assumes azimuthal symmetry of both the dust clouds (on 10s of pc scales) and the differential scattering cross-section, the latter of which only holds true for spherical particles. For non-spherical particles, employment of anomalous diffraction theory demonstrates that the resulting dust scattering halo image can range from ellipsoidal to a nearly diamond shaped pattern, depending on the elongation and relative degrees of alignment among the dust grain population, induced by the presence of a magnetic field in the ISM \citep{Draine2006}.

\textbf{Time variable scattering} \\ 
In the majority of cases, dust scattering halos are treated as static -- a common approximation that reduces the complexity of the problem. However, in reality, dust scattering halo images are time variable because there is a path-length difference between the  scattered light and non-scattered light. The image arising from a steady, unchanging point source can be referred to as a ``static'' or ``quiescent'' dust scattering halo. If the source of X-ray light undergoes a rapid high fluence outburst,  ring images will be produced as the light from the flare propagates through interstellar clouds (Fig.~\ref{fig:HaloGeometry}). The surface of equal time-delay as the X-ray wavefront propagates through the interstellar medium is an ellipse, and each dust cloud intersecting that ellipse produces a separate ring. As the ellipsoidal surface grows with time, the angular sizes of the ring echoes also increase with time.
This phenomenon is often referred to as a dust scattering echo or dust ring echo. 
The first dust scattering echoes were observed as a result of the X-ray afterglows from extragalactic gamma ray bursts, which scattered off of local Galactic dust, producing ring images that were resolvable by both \xmm\ \citep{Vaughan2004, Tiengo2006} and the Neil Gehrels Swift Observatory \citep{Vaughan2006, Vianello2007}.

Dust scattering rings can be used to (i) measure the distance to the X-ray source, given the line-of-sight distribution of dust and knowledge of the X-ray light curve \citep{Heinz2015}, (ii) measure the line-of-sight dust abundance (ISM tomography), given full knowledge of the distance and X-ray light curve of the background source \citep{Heinz2016}, or (iii) estimate the time and fluence of an X-ray burst, given knowledge of the line-of-sight dust position and abundance. 
The interplay among all these parameters is described by the time delay ($\delta t$) associated with a particular observation angle ($\theta_{\rm obs}$, the angular distance between the central point source and the observed scattering image) \citep{Trumper1973}:
$$ \delta t = \frac{D}{2c} \frac{x \theta_{\rm obs}^2}{(1-x)}, $$
using the geometry described in Fig.~\ref{fig:HaloGeometry}. 
In the case of ring echoes originating from extragalactic sources, $x << 1$, so the distance to the dust clouds can be measured directly from the angular size of the ring echoes. 
Following from the dust scattering halo integral above, the surface brightness profile for a dust ring echo is
$$ \frac{dI}{d\Omega}(E, \theta_{\rm obs}, t) = \int_{x_{\rm min}}^{x_{\rm max}}\int_{a_{\rm min}}^{a_{\rm max}} F_a(E, t-\delta t)\ \frac{d\sigma_{\rm sca}}{d\Omega}(E, a, \theta_{\rm sca})\ \frac{n_d(a)}{(x_{\rm max} - x_{\rm min})}\ da\ dx,$$
where $\delta t$ and $\theta_{\rm sca}$ are a function of $x$, and it is assumed that the dust cloud with an abundance and grain size distribution described by $n_d$ is uniformly distributed between $x_{\rm min}$ and $x_{\rm max}$. 
Even slow changes in an X-ray light curve can lead to measurable differences in a dust scattering halo surface brightness profile as a function of time, which can be used to constrain the distance to the X-ray emitting source (or inversely, the position of the dust). The first distance measurement obtained in this way was for Cyg~X-3 \citep{Predehl2000}. 
Data-driven methods of studying distances and line-of-sight positions of dust clouds include cross-correlation between light curves from an annulus centered on $\theta_{\rm obs}$ and the light curve of a central point source \citep{Ling2009}, and de-convolution of the dust scattering halo image as a function of time using the predicted scattering halo intensity profile as a kernel \citep{Heinz2016}. Both methods can potentially be used to study dust scattering halo variability at a lower contrast than that typically seen for ring echoes.

\subsection{The X-ray fine structure}\label{par:xafs}

The photoelectric effect describes the interaction of an incoming photon and the electrons in the atom. As the energy of the photon equals the binding energy of the electron, it is ejected from the atom with kinetic energy equal to $E-E_0$, where $E$ is the incoming photon energy and $E_0$ is the electron binding energy. The X-ray photon energy is sufficient to remove electrons from the innermost levels (K, L, M). The resulting spectroscopic feature displays first a sharp drop at the energy corresponding to the electron binding energy, then the probability of the effect to take place decreases exponentially, as the material becomes more transparent to photons with energy larger than the threshold one, creating a characteristic saw-tooth feature.

\begin{figure}
\begin{center}
\includegraphics[angle=0,keepaspectratio=true,scale=0.4]{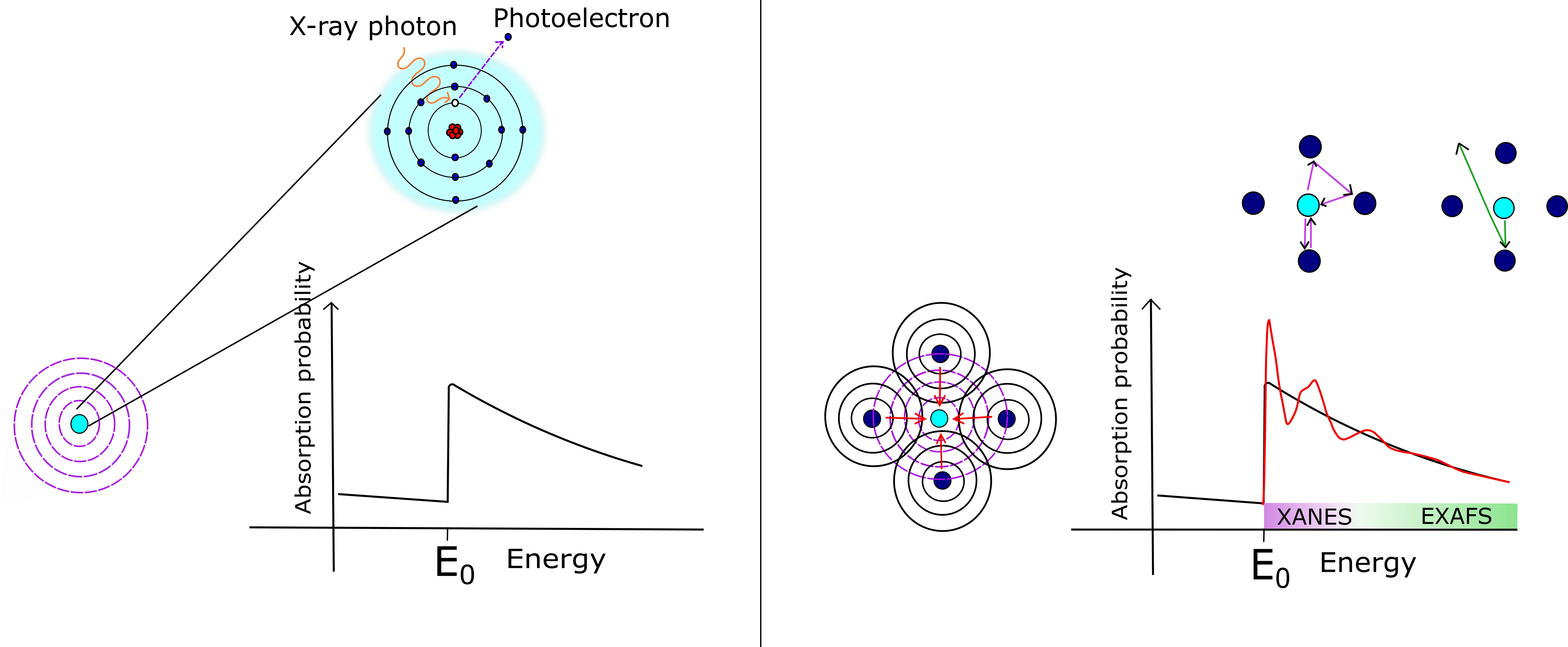}
\caption{\label{fig:xafs}{\it Left panel}: in a gas, the incoming X-ray photon will remove one of the inner orbit electrons to the continuum, without further interaction. The resulting cross section has a saw tooth shape, starting at $E_0$, the binding energy of the removed electron. {\it Right panel}: In a solid, the photoelectron wave of the removed electron interacts with the ones of the neighboring atoms, adding a characteristic interference pattern to the photoelectric cross section. The regime of features near $E_0$ (pink shade and arrows in the insert) is called XANES, due to multiple scattering among the atoms, while the structures farther from $E_0$ are dominated by single scattering (EXAFS, green shade and arrows in the insert). Figure adapted from \citep{zeegers_thesis}.}
\end{center}

\end{figure}

{\textbf{Absorption fine structure}} \\ 
If the X-ray photons interact with solid particles, rather than gas, additional effects take place, creating the spectral features of EXAFS (extended X-ray absorption features) and XANES (X-ray absorption near edge structures), called collectively XAFS (X-ray absorption fine structure). The basic mechanism is illustrated in Fig.~\ref{fig:xafs}. The incoming photon interacts with the electron in the shell. The electron can be described as a wave that interacts with the neighbouring ones, creating positive and negative interference. This diffraction pattern depends on the number of electrons and their distance to the nucleus, therefore revealing the chemical bonds in the grain. XANES appear as sharp features in the vicinity of the threshold energy  and they are the result of multiple backscattering from the neighboring atoms, while EXAFS, that are lower amplitude and broadened features, are visible at energies above 50--100\,eV the threshold energy. They are the result of a single backscatter from the neighboring atoms.

\textbf{Scattering fine structure} \\ 
By the fundamental nature of dielectrics, an absorption resonance  also yields a scattering resonance. As a consequence, the Rayleigh-Gans approximation also breaks down near the $n=1$ (K shell) and $n=2$ (L shell) photoabsorption features that are used to probe the metals comprising interstellar dust. More exact calculations for the scattering cross-section, via either anomalous diffraction or Mie theory, demonstrate that every K and L shell absorption resonance has a complementary decrement in scattering efficiency \citep{Martin1970}. This X-ray scattering fine structure (XSFS) signature affects the spectral shape of dust scattering halos as well as the apparent shape of photoabsorption, due to the contribution of dust scattering to the total extinction through the ISM \citep{Hoffman2016, Corrales2016}. This fact makes it so that XSFS can also be investigated through direct high resolution spectroscopy of point sources (Sect.~\ref{par:mie} and Fig.~\ref{f:size}).

\begin{figure}
\begin{center}
\includegraphics[width=\textwidth]{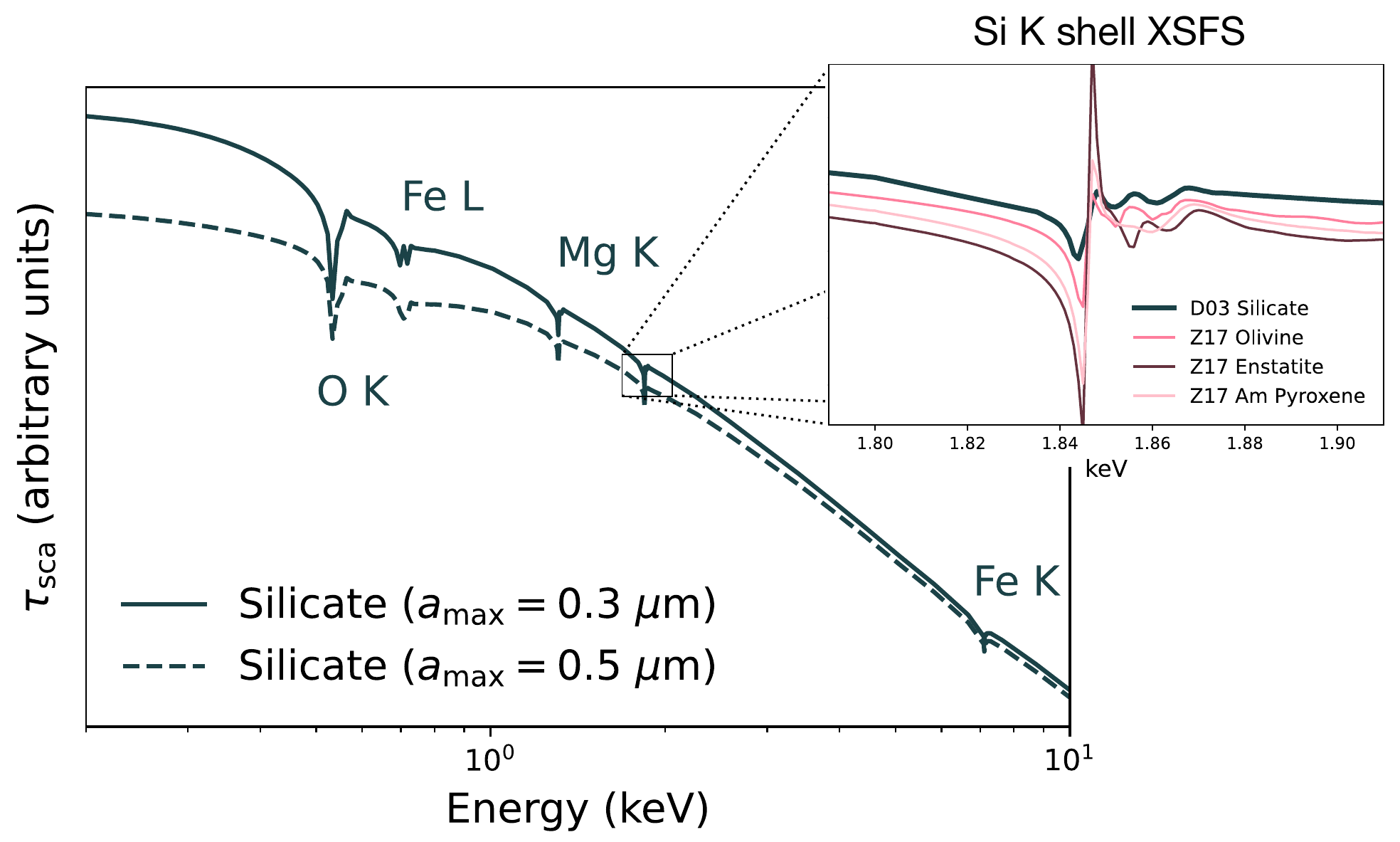}
\caption{
A demonstration of X-ray scattering fine structure (XSFS) computed for a power law distribution of spherical dust grains, using the optical properties for silicates produced by \citet[][]{Draine2003}. Extending the MRN distribution to larger grains (dashed line) causes a departure from the $E^{-2}$ dependence, characteristic of Rayleigh-Gans scattering, at a higher energy compared to the standard MRN distribution (solid line). The normalization for each curve has been scaled for ease of comparison. 
The inset shows XSFS around Si K in more detail. The laboratory derived optical constants from \citet[][]{Zeegers2017} were used to compute the XSFS for several astrosilicate candidate materials.
}
\label{fig:xsfs}
\end{center}
\end{figure}

A dust scattering halo spectrum, when compared to the central point source, provides a direct measurement of the scattering opacity from dust in the ISM via the equation
\begin{equation}
    \tau_{\rm sca} = \ln \left( \frac{F_h}{F_{ps}} + 1 \right)
\end{equation}
where $F_{h}$ is the spectrum of the halo and $F_{ps}$ is the point source spectrum after the effects of all sources of line-of-sight extinction.  
Using this method, the first detection of XSFS from a dust scattering halo was that around the low mass X-ray binary Cyg~X-2, where the strength of the O K shell resonance is strong enough to become apparent in low resolution spectra \citep{Costantini2005}. With sufficient resolution, XSFS may be used to discern the mineralogy of dust (Figure~\ref{fig:xsfs}, inset). 
As with dust scattering halos, the exact profile of XSFS depends on a variety of factors including grain size distributions, shape, and alignment \citep{Hoffman2016}. The contribution of XSFS to apparent extinction is also dependent on the shape of the complementary scattering halo surface brightness profile and the spatial resolution of the X-ray spectrograph \citep{Corrales2016}. Finally, the broad-band spectral energy distribution of a dust scattering halo is also affected by the dust grain size distribution. 
In the example shown in Figure~\ref{fig:xsfs}, extending the MRN distribution to a grain size of $0.5~\mu{\rm m}$ enhances scattering and changes the shape of the dust scattering opacity curve so that it flattens at higher energies than the typical MRN distribution.

\subsection{Correcting X-ray Observations for ISM Attenuation}\label{par:scattcorrection}

\begin{figure} 
    \includegraphics[width=\textwidth]{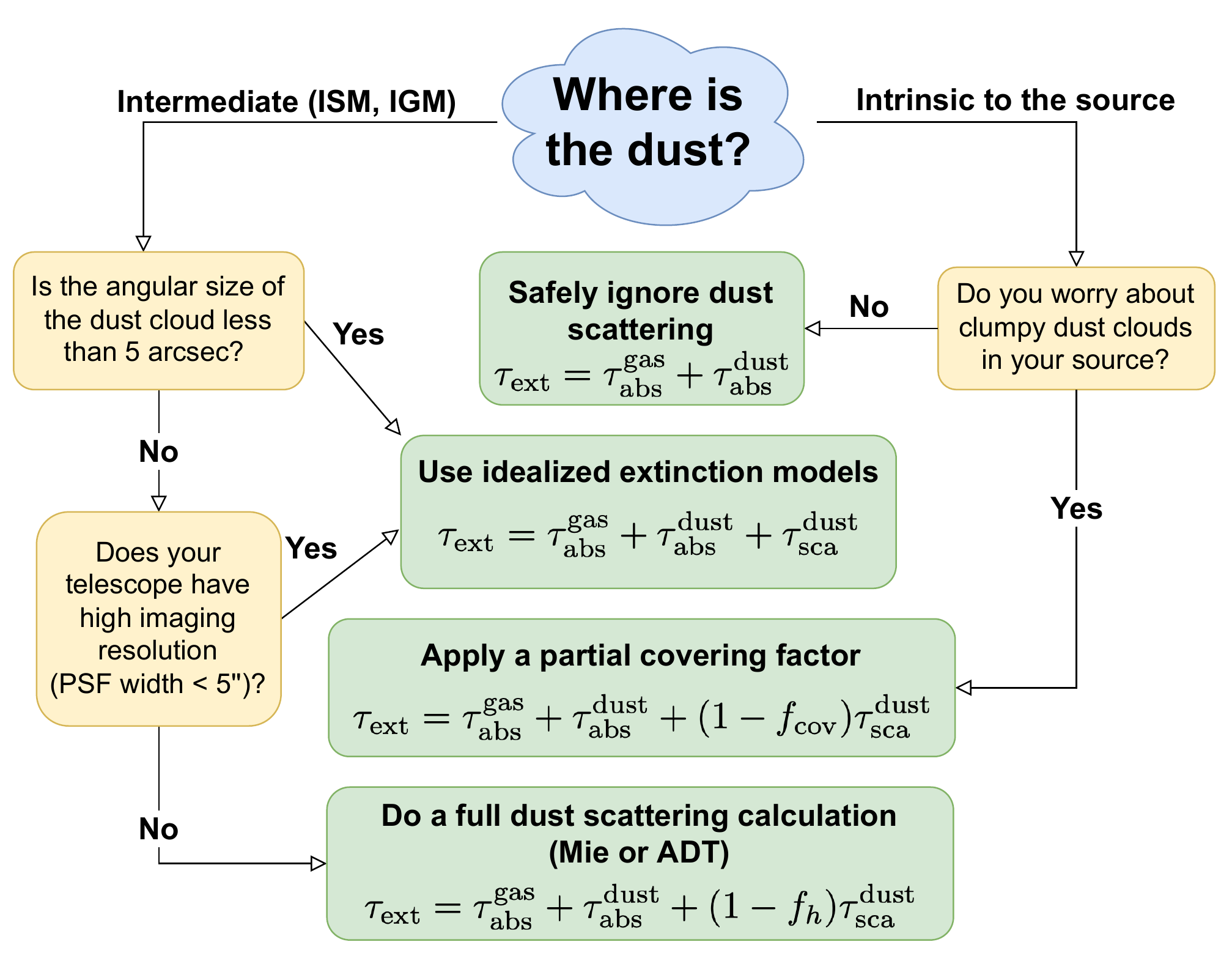}
    \caption{
    Decision tree describing when and how to include dust scattering when fitting spectral models for a point source obscured by cool gas and dust, based on the scenarios described in \citet{Corrales2016}. In each spectral model, the attenuation by gas and dust is broken down into three component: the opacity from gas absorption ($\tau_{\rm abs}^{\rm gas}$), the absorption component from dust ($\tau_{\rm abs}^{\rm dust}$), and the dust scattering optical depth ($\tau_{\rm sca}^{\rm dust}$). 
    Dust that is intrinsic to the X-ray emitting system, but does not fully cover the X-ray source, contributes a fraction of the total dust scattering opacity towards extinction, according to the covering factor $f_{\rm cov}$. 
    Dust that is intermediate along the line of sight may require a full dust scattering halo calculation if the telescope imaging resolution is worse than $5-10''$. In this case, the user will need to determine what fraction of the scattering halo is enclosed by the image region used to extract a spectrum ($f_h$). 
    Any modelers that wish to examine the effects of non-standard (e.g., MRN) dust grain size distributions that include a contribution from large $> 0.3~\mu{\rm m}$ grains will need to do a full scattering halo calculation.
    }
    \label{fig:DecisionTree}
\end{figure}

When correcting X-ray observations for attenuation by the foreground ISM, it is important to separate the contributions from the gas phase (pure absorption) and the solid phase (dust absorption and scattering). It has been demonstrated that not including dust scattering in X-ray spectral models can yield different conclusions for the physical properties derived for the underlying source emission, such as the disk black-body temperatures in X-ray binaries, on the order of $30-50\%$ \citep{Smith2016}. The modeling techniques for dust attenuation will depend on geometrical effects, including the relative distances between the dust and X-ray source, grain size distribution, and imaging resolution of the spectroscopic instrument \citep{Corrales2016}. Figure~\ref{fig:DecisionTree} describes the decision making process for including dust scattering effects when evaluating an astronomical spectrum of interest.

\section{Laboratory measurements of solid particles}

In order to interpret X-ray astronomical data, a comparison with reliable dust models is necessary. Thanks to experimental measurements campaigns, targeted specifically at analogues of interstellar dust materials, absorption profiles have been obtained of a number of X-ray edges: Fe\,L \citep{lee09,westphal19}, O\,K \citep{psaradaki20}, Al\,K\citep{costantini19}, Mg\,K \citep{rogantini19}, Si\,K \citep{Zeegers2017,zeegers19} and Fe\,K \citep{lee05,rogantini18}\\ Depending on the edge energy and the measured sample thickness, different facilities and techniques are necessary in order to obtain the absorption profile. For the works cited here, the measurements were performed using synchrotron beamlines (with transmission and fluorescence techniques) as well as electron energy loss  spectroscopy (EELS).\\

{\bf Transmission}\\
A polychromatic synchrotron light beam is converted, through a monochromator, into a mono-chromatic energy beam ($I_0$) that interacts with the specimen. The outcoming radiation ($I$) is attenuated according to the Beer-Lambert law: 

\begin{equation}\label{eq:bl}
I=I_0 exp(-t\mu),
\end{equation}
where $t$ is the thickness of the material and $\mu$ is the absorption coefficient, which depends on the atomic- ($Z$) and mass number ($A$) of the sample as well as the incident energy ($E$) and the density of the sample ($\rho$) according to $\mu\approx\frac{\rho Z^4}{AE^3}$ \citep[][]{newville14}.\\

{\bf Fluorescence}\\
With this technique, the secondary effect of the photoelectric effect of fluorescence is measured \citep[][]{newville14}. As fluorescence photons are emitted at specific energies, this information can be used to recover the absorption profile. The absorption coefficient $\mu(E)$ can be approximated as $I/I_F$, where $I$ is again the incident intensity and $I_F$ is the measured intensity of the fluorescent emission.\\

{\bf Electron energy loss spectroscopy}\\
With this technique, the specimen is targeted by a beam of monochromatic electrons. The Coulomb interaction depend on whether the electrons interact with the nucleus, with peripheral electrons or with tightly bound electrons. In the last case an inelastic scattering event occur, which implies an exchange of energy. The incoming electron beam then lose energy depending on the binding energy of the atoms in the target, producing a spectral energy distribution, characteristic of the material. The outcoming kinetic energy of the electrons mirrors the absorbing part of the dielectric function \citep[][]{egerton11}. \\

The attenuation of a beam through a medium can be described by Eq~\ref{eq:bl}. As the absorption profile of the material has been extracted, it will present as in Fig.~\ref{fig:xafs} (right panel), with XANES features near the edge and EXAFS features at $E>50-100$\,eV from the inset of the edge. 
The latter can be expressed in analytical terms \citep[e.g., ][]{teo86}. The function $\chi$ represents the superimposing sub-structures to the photoelectric edge shape:

\begin{equation}
\chi(E)=\frac{\mu(E)-\mu_0(E)}{\Delta\mu}
\end{equation}

where $\mu_0$ is the absorption coefficient of the smooth underlying continuum \citep[][]{newville14}. The function $\chi(E)$ is normalized by the edge jump at $E_0$ (Fig.~\ref{fig:xafs}), through the term $\Delta\mu$. The technique to analyse EXAFS prescribes first a representation of $\chi$ into the wave number space $k=\sqrt{\frac{2m(E-E_0)}{\hbar^2}}$. Then, using the fact that the frequency of the EXAFS features depend on the distance of the absorbing atom and the neighbouring scattering electrons, a Fourier transform of $\chi(k)$ would show a photoelectron scattering profile as a function of the distance. The detection of reliable EXAFS, which are shallow and broadened spectral features, is however strongly dependent on data quality. While at a synchrotron facility a detection may be achieved, in an astronomical context, EXAFS are not regularly observed. Signal-to-noise ratio of the data is the main limiting factor. In addition, a spectrum of an astronomical source may display many different features, from both intrinsic and intervening gas that 
would blend with the EXAFS, confusing the extraction of the signal. On the contrary, XANES, thanks to their sharpness and relatively high amplitude, can be well detected also in an astronomical X-ray spectrum (Fig.~\ref{fig:xafs}). The technique to treat the multiple scattering process in a material is not straightforward as for EXAFS, and has been limited in the past by the heavy calculation required in following the scattering paths \citep[][for a review]{rehr00}. Calculations are implemented in {\it ab-initio} codes, e.g., FEFF \citep[][]{feff98} or Quantum Espresso \citep[][]{quantum09}, among others, that provide comparable results with experimental data \citep[e.g.,][]{takahashi18}. 

\subsection{Implementation to astrophysical models}\label{par:f1f2}

Once the intensity profile of a material has been obtained, it needs to be post-processed to be adapted to be part of an astrophysical model. Here we consider the case of a specimen observed in fluorescence by the LUCIA-beamline at the Soleil synchrotron facility. This description closely follows the procedure adopted in \citet{Zeegers2017,zeegers19,rogantini19} and \citet[][]{costantini19}. Some instrumental effects should be first taken into account. For example, pile-up that occurs when two or more photons are recorded at the same time, resulting as one photon with double the energy. The result is a spurious line feature that can be easily corrected for. If the sample is  sufficiently thick, self absorption in the sample will cause the fluorescence intensity to be visibly attenuated \citep[][]{Zeegers2017}. Several empirical methods can be used to correct for this effect \citep[][]{stern95, booth05}.\\ 
Usually, multiple fluorescence observations are performed on the same specimen. If the data quality allows it, these are added to increase the signal-to-noise ratio of the measurement. These data are then inverted to mock a transmission profile. The pre- and post edge are then fitted to a theoretical profile of known thickness (typically $\tau=0.01\mu$m). The values tabulated starting from \citet{henke93}, provide an excellent resource to compute, in first approximation, a transmission profile of a given material of known composition, density and thickness.

The obtained profile is related to the imaginary part $k$ of the refraction index $m=n+ik$. The optical constants $n$ and $k$ are unique signatures of a material. 
Sometimes they are expressed with a different notation e.g., as dielectric functions, $\epsilon_1$ and $\epsilon_2$ \citep{Draine2003}, or as atomic scattering factors $f_1$ and $f_2$. If the incident radiation has a wavelength $\lambda$ larger than atomic dimension, or if the scattering angle is small, $f_1$ and $f_2$ are then not dependent from the scattering angle \citep{henke93}. Optical constants and scattering factors can be easily transformed into one another \citep[see][]{rogantini18}:

\begin{equation}
n(E)=1-\frac{\rho N_A r_0}{2\pi A}\lambda^2f_1(E),    
\end{equation}
\begin{equation}
k(E)=\frac{\rho N_A r_0}{2\pi A}\lambda^2f_2(E).
\end{equation}

Here $\rho$ is the density of the material, $N_A$ the Avogadro number, $r_0$ the electron radius, $A$ the atomic mass and $\lambda$ the wavelength of the incoming radiation. 
The atomic scattering factors depend on one another according to the Kramers-Kroning relations \citep{kramers26,kronig26}. Therefore the real part of the refraction index can be derived \citep{watts14,henke93}:\\ 
\begin{equation}\label{eq:f}
f_1(E)=Z^\star -\frac{2}{\pi}\int_0^\infty\frac{\epsilon f_2(\epsilon)}{{{\epsilon}}^2-E^2}d\epsilon,
\end{equation}
where $Z^\star$ can be approximated as:

\begin{equation}
Z^\star\approx Z-(Z/82.5)^{2.37}.
\end{equation}
The factor $Z^\star$ is a small relativistic modification of $Z$, in a high-energy photon limit \citep{cromer70}. This reduction in $Z$ is only relevant for high-$Z$ elements. As observed for instance in \citet{henke93}, Eq.~\ref{eq:f} displays a discontinuity when $\epsilon=E$. Furthermore, a true atomic scattering factor needs in principle to be defined at all energies, while experimental energy ranges are in fact limited. Recently, the use of a piecewise Laurent polynomial algorithm proved to offer an accurate description near the edge \citep{watts14}, avoiding the requirement of an infinite coverage in energy or an homogeneous binning of the data. The real and imaginary parts of the refractive index are a necessary input to calculate the scattering and absorption efficiency in the interaction between the incoming X-rays and the dust particles.  

\subsection{Interaction of X-rays with dust grains}\label{par:mie}
The interaction of a set of plane parallel electromagnetic waves with a spherical particle is described by the Mie theory \citep{mie}, when the 
size parameter $X=2\pi a/\lambda$ does not exceed $2\times10^4$ \citep{BohrenHuffman}. In this relation, $\lambda$ is the wavelength of the incident radiation and $a$ the radius of the scattering sphere. The spherical shape of the target allows to separate the solution into the radial and angular dependence, in a form of infinite series of spherical multipole partial waves. The solutions are the scattering ($Q_s$) and extinction efficiency ($Q_e$). The absorption efficiency is simply given by: $Q_a=Q_e-Q_s$. The relative cross sections relate to the efficiency through the size of the particle: $C=Q\pi a^2$. 

Conditions where the size parameter $X$ was much larger were not initially foreseen in Mie theory codes \citep[e.g.,][]{wiscombe}. However, a large value of $X$ can occur in various astrophysical contexts \citep[and references therein]{wolf}. Subsequent implementations allowed to include arbitrarily large values of $X$ \citep[MieX,][]{wolf}, either to be able to model scattering from very large grains or take into account very short wavelengths.\\ 
When the particle size is much larger than the wavelength, the van der Hulst's Anomalous Diffraction Theory \citep[ADT,][]{vandeHulst} provides an analytical approximation:
\begin{equation}
Q_s=2-\frac{4}{p}sin(p)+\frac{4}{p^2}(1-cos(p)),
\end{equation}
where $p=4\pi a (\tilde{n}-1)/\lambda$ and $\tilde{n}$ is the ratio between the refractive index inside and outside the sphere. This approximation is only valid if $\tilde{n}$ does not deviate significantly from unity, indicating that the refractive index in the sphere is not very different from the ambient space, causing only a small shift of the out-coming wave. 
In the X-ray range, both the Mie theory (especially the latest implementations) and the ADT can be used. In particular \citet{Draine2006} showed that for $E\geq 60$\,eV and $a>0.035\mu{\rm m} (\frac{60 {\rm eV}}{E})$, silicate grains can be safely treated using the ADT approximation.

In the ISM, grains are not spherical, but rather porous and elongated. The effects of geometries different than spherical (e.g., spheroids) have been explored in \citet{Hoffman2016}, using an extension of ADT \citep[GGADT, ][]{ggadt}. Different types of disordered aggregates (ballistic aggregates, BA) mimic different degrees of porosity, as defined in \citet{shen08}. While the extinction cross section for spheroids does not undergo significant changes with respect to spheres, BA may show an increased absorption cross section near the edge features \citep[][]{Hoffman2016}. This effect appears more important for edges at low ($E<1.3$\,keV) energies and become less visible for e.g. the Si\,K and Fe\,K edges.

\begin{figure} 
\hspace{-0.7cm}
    \includegraphics[width=1.1\textwidth]{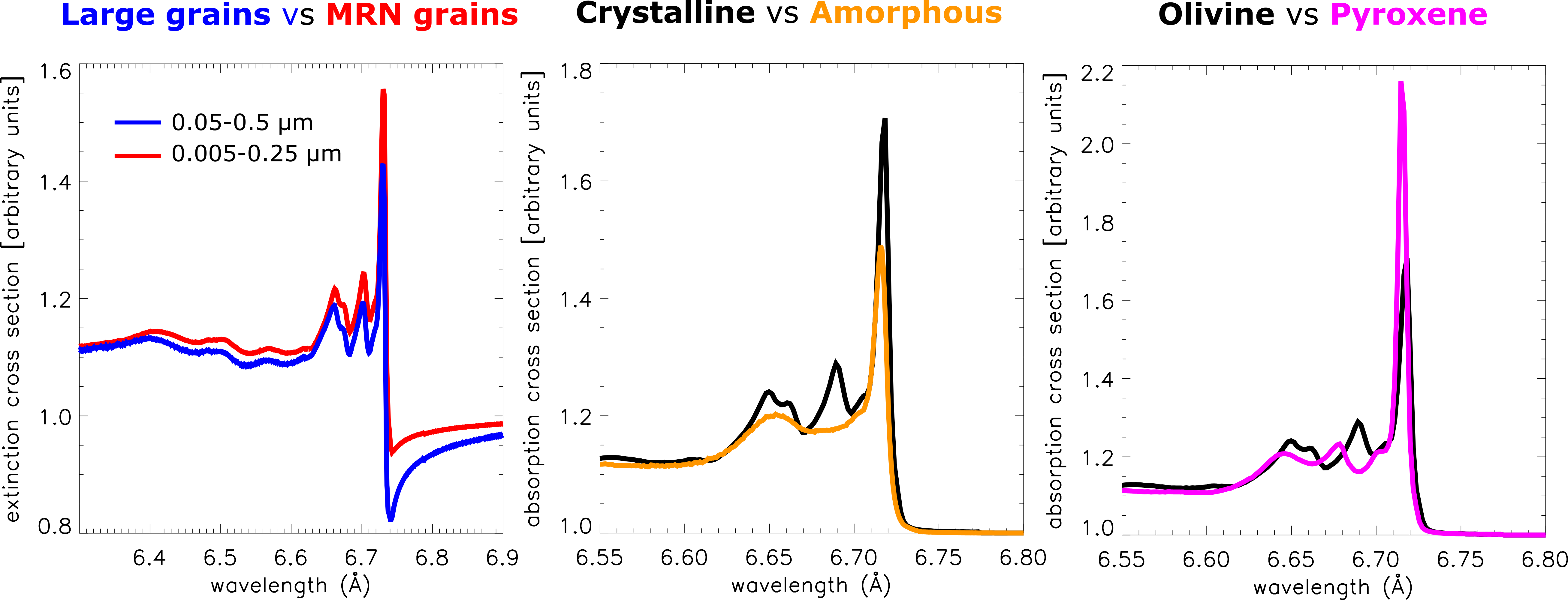}
    \caption{Impact of different grain properties on the shape of the absorption or extinction cross section. Left panel: the effect of large grains on the extinction cross section is to enhance the scattering peak. Here an MRN distribution (red line) is compared with a distribution shifted towards larger grains (blue line). Center Panel: the degree of crystallinity in a grain may be visible in the absorption cross section. The disordered organization in the lattice causes the absorption features to be smeared out (yellow line). Right panel: different electron configurations entail a different XAFS pattern (Sect.~\ref{par:xafs}). Here olivine (MgFeSiO$_4$, black line) is compared with a pyroxene (Mg$_{0.75}$Fe$_{0.25}$SiO$_3$, purple line). Courtesy of S. Zeegers.}
    \label{f:size}
\end{figure}

The extinction cross section can be calculated for a range of energies and grain radius, via the size parameter $X$. As in the ISM X-rays interact with a variety of grain sizes, rather than a single one, a grain size distribution should be applied (Sect.~\ref{s:size}). In Fig.~\ref{f:size} (left panel) a comparison between the MRN size distribution and a distribution with sizes ranging in the interval 0.05--0.5\,$\mu$m is shown. In this example the large grain distribution follows the same powerlaw shape as prescribed in the MRN model. The modulation of the region at the longer wavelength side of $E_0$ (XFSF, Sect.~\ref{par:xafs}), due to the scattering term $n$ in the refraction index, is therefore sensitive to the grain size distribution, while the absorption pattern, at shorter wavelengths, is not significantly affected in shape. Absorption however is slightly less efficient as a function of grain size, as can be seen in the post edge of Fig.~\ref{f:size} (left panel).
When modeling the transmission of a smooth continuum source through interstellar dust, the characteristic dip in scattering efficiency near a photoabsorption feature can mimic the appearance of an emission feature, while it may be in fact the sign of large grains contribution to the extinction \citep[][]{Zeegers2017,rogantini19}.

The central panel of Fig.~\ref{f:size} shows the difference in the profile of the cross section between an amorphous and a crystalline material (olivine in this case). The XAFS of the amorphous material will appear smoother, depending on the degree of amorphisation of the material \citep{zeegers19}. This smoothing is due to the disordered organization of the atoms in a glassy material. 
The XAFS near the absorption edges can provide direct information on the chemistry of the intervening matter (Fig.~\ref{f:size}, right panel). Here the absorption profile of olivine (FeMgSiO$_4$) is compared with a pyroxene (Mg$_{0.75}$Fe$_{0.25}$SiO$_3$). Both the amplitude and the position of the XAFS serve as a chemical composition diagnostics \citep[see][for more examples in the silicon region]{zeegers19}.

Finally, once the extinction cross section has been calculated, it can be implemented in any spectral fitting program, such as XSPEC \citep{arnaud} and SPEX \citep{kaastra96}. 

\section{Scattering and absorption of X-rays: the state of art}\label{par:stateofart}
Ever since the scientific investigation of the ISM began, gas and dust have been observed at all wavelengths. The study of ISM in the X-ray band has been hampered in the past by instrumental limitations, while at longer wavelengths a significantly better energy and spatial resolution allowed a deep understanding of the ISM properties. The X-ray band however offers several advantage points. In particular:

{\it (i)} The broad band energy coverage (0.1--10\,keV) of present X-ray observatories encompasses a variety of transitions, from neutral to highly ionized gas, of the fundamental metals in the Universe: C, N, O, Ne and Fe, among others. On the dust observation side, the X-ray band covers all the features pertaining to the major dust constituents. The photoelectric edges of neutral C, O, Mg, Si and Fe fall in the X-ray band. While other elements can be relatively easily investigated, carbon can be at the moment only reached by the LETG spectrometer on board of {\it Chandra}. This spectral region however is complicated by instrumental effects that make the study of the astronomical carbon edge challenging \citep[e.g.,][]{schmitt}.

{\it (ii)} Transmission spectra in X-rays can be obtained from a large range of column densities (log$N_{\rm H}({\rm cm}^{-2})=20-23$). Provided a bright X-ray source behind an ISM layer, the large penetrating power of X-rays allows the investigation of a variety of environments of our Galaxy, from the tenuous diffuse medium to molecular clouds (see Fig.~\ref{fig:nh}). The diffuse medium, from a X-ray point of view, includes  a range of column densities, log$N_{\rm H}({\rm cm}^{-2})$, of about 20--21.7. In this regime, the oxygen K and iron L edges are well visible at low energies. As the column density increases, (log$N_{\rm H}({\rm cm}^{-2})\sim$21.7--22.7) the soft X-ray edges become more absorbed and they are eventually lost into instrumental noise. At the same time, the Mg and Si\,K edges become prominent. At these column densities a variety of environments can be sampled: from extended or low density molecular clouds to far-away lines of sight that cross more than one Galactic arm, for example towards the Galactic center. Finally, X-rays can also access highly absorbed lines of sight (log$N_{\rm H}({\rm cm}^{-2})\sim$23), corresponding to $A_V\sim$50. At this moment however, the deep iron K edge at 7.1\,keV (Fig.~\ref{fig:nh}) cannot be characterized, due to the still moderate energy resolution and sensitivity of the current instruments in this spectral region \citep[e.g.,][]{rogantini18}.

{\it (iii)} The gas and the dust components of the cold phase of the ISM can be studied along the same line of sight, giving a direct information on depletion and abundance of a given element. This is especially true for oxygen, whose \ion{O}{I} $1s-2p$ prominent transition at $\sim$23.5\,\AA, lies next to the O\,K photoelectric edge at $\sim$23.3\,\AA \citep{gatuzz14,psaradaki20}. For other elements included in both gas and dust, modeling can be more complicated, as the gas transitions may lie in the same energy range of the edge structure \citep[e.g. the iron L edges at 17.1\,\AA,][]{costantini12}. In some situations, there may be no very prominent gas transitions \citep[e.g. near the Si edge at 6.74\,\AA,][]{rogantini19}.

{\it (iv)} The two components of extinction, i.e. scattering and absorption, may be investigated along the same line of sight. Both imaging and spectroscopy can indeed provide information on the dust chemistry, grain size distribution, and dust clumpiness along the line of sight \citep{Corrales2016}.

{\it (v)} Finally, many of the bright X-ray sources in the Galaxy are located along the plane \citep[$\vert b\vert <12$\,deg,][]{Predehl1995}. This provides a very effective sampling along different lines of sight where most of the cold phase of the ISM resides. Low-mass X-ray binaries, thanks to their almost featureless intrinsic broad band spectrum, are the sources most suitable for dust studies. High-mass X-ray binaries and supernova remnants, display a spectrum rich of emission lines over a broad energy interval. This hampers, in general, the detection of the dust fine structure close to the absorption edges. \\

Dust modeling from the X-ray point of view aimed at first at confirming the findings obtained at long wavelengths. However, given the complementary and the advantages provided by the X-ray properties of the ISM, results have also challenged the common knowledge built up so far. Here we try to summarize our current understanding of dust from the X-ray side. \\

\noindent
{\bf Chemical composition} \\
The strength of the X-ray band is to display, for a given column density, at least two visible absorption edges in an absorbed spectrum, potentially belonging to silicates. This allowed the study of these materials along lines of sight with different column densities.  The sample of sources, sometimes revisited with different instruments (namely, the \xmm-RGS, \ch-LETG and \ch-HETG grating spectrometers), are in general X-ray-bright low-mass X-ray binaries ($Flux(2-10\,{\rm keV}) > 10^{-9}$\,erg\,cm$^{-2}$\,s$^{-1}$).\\  
From the study of both the individual edges and a simultaneous modeling of the O\,K and Fe\,L edges, prominent in the diffuse medium, it has been reported the presence of Mg-rich silicates both in absorption \citep[][]{costantini12,psaradaki20,valencic13} and from scattering halos \citep{Costantini2005}. Amorphous olivine does not seem to play a major role towards those studied lines of sight. \citet[][]{valencic13} report for example a ratio of enstatite over olivine of about 3.4. As seen in Sect.~\ref{par:star}, Mg-rich silicates would be consistent with a chain of dust condensation events in equilibrium conditions in a stellar wind. 
However, even under these conditions amorphous olivine should be present in significant amount. The GEMS-like particles, amorphous, Mg-rich silicates, would be consistent with the X-ray results. In a dedicated experiment, however, the iron-L edge of one sight-line has been compared with real GEMS, returned by the Stardust mission from the comet 81P/Wild 2 environment \citep[][]{westphal19}. The iron profile of GEMS turned out to be incompatible with the astronomical data of the ISM, indicating that cometary fine-grained material might not be a proxy for interstellar dust.

The iron edges provide in principle a direct view of iron in the ISM. Given the extreme depletion of this element, the edge shape is mostly determined by dust absorption. Along diffuse lines of sight, the iron L-edge has been found to be dominated either by oxides (e.g., Fe$_2$O$_3$) in a mixture of different silicates \citep{lee09}, or metallic iron \citep{costantini12} or FeS mixed with metallic iron \citep{westphal19}. It must be noted however that different sets of dust models were used in different works, especially for iron compounds. Therefore, this apparent discordance in results may be still attributed to a difference in modeling and completeness of the data bases.

\begin{figure}
    \centering
    \includegraphics[width=0.7\textwidth]{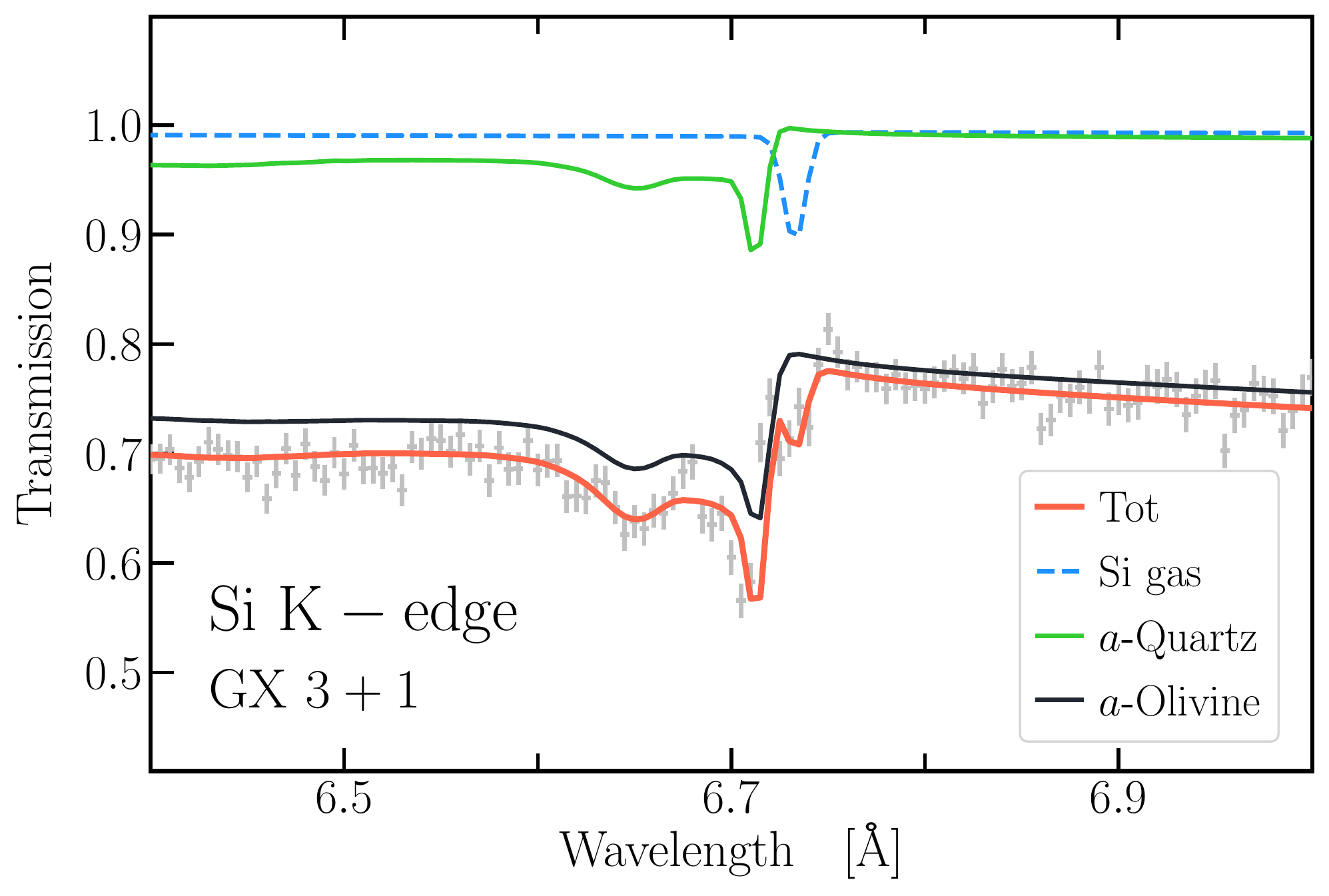}
    \caption{\label{f:gx3p1} \ch-HETG data around the Si edge along the line of sight of GX~3+1. The best fit (red line) is dominated by amorphous olivine (black line). Minor amounts of other types of dust and neutral gas were also detected. The data are normalized for the continuum emission intrinsic to the source. Adapted from \citep{rogantini20}.}
    
\end{figure}

Lines of sight characterized by a higher column density (log$N_{\rm H}({\rm cm}^{-2})\sim$22.1-22.9), have been more consistently studied in the recent literature, using the same set of laboratory measurements for Mg and Si. The modeling of the Si\,K edge alone \citep[][]{zeegers19} and the combined Mg and Si\,K edges \citep[][]{rogantini20} pointed to a different scenario with respect to the more diffuse medium. The contribution of amorphous olivine has been reported to be dominant along these lines of sight. 
Considering the high signal-to-noise ratio data in the samples of source in \citet[][around 10 sources]{zeegers19,rogantini20}, olivine has been deemed to contribute more than 60\%, and up to 80\%, to the total dust budget for a given source. In Fig.~\ref{f:gx3p1} an example of Si absorption along the line of sight of GX~3+1 is given. The majority of absorption is attributed to amorphous olivine in this fit \citep{rogantini20}. Although the presence of olivine has been determined with high confidence, it has to be noted that for relatively large values of $N_{\rm H}$ the iron L edges are inaccessible. Therefore the simultaneous presence of any metallic iron, iron sulfide or iron oxides, could not be tested. \\ 

\noindent
{\bf Dust size distribution} \\
The MRN dust size distribution has been often preferred in the modeling, by virtue of its simple analytical form (Sect.~\ref{s:size}). By means of high-resolution X-ray spectroscopy, indications of a deviation from this distribution have been sometimes reported. \citet[][]{westphal19} report a lower limit on the size of larger grain contribution of $a >245$\,nm along the line of sight of Cyg~X-1. \citet[][]{Zeegers2017} report an improvement of the fit towards  GX~5-1 if a distribution with a size range of 0.05--0.5$\mu$m is adopted. Although the scattering feature in the pre-edge region can be a predictive tool for the dust distribution (Fig.~\ref{f:size}), in practice, its use may be hampered by several factors. Low signal to noise in spectral regions that may be crowded with lines from the ISM may be a contributing factor \citep[e.g.][]{psaradaki20}. Sometimes instrumental features may also confuse the picture \citep[e.g.][]{rogantini20}. Finally, the scattering peak strength is energy dependent \citep[][]{Hoffman2016,Draine2003}, therefore some edges may not provide useful information.

Results from scattering halos indicate that the dust size distribution may not be homogeneous from cloud to cloud. This can be quantified especially when bright and well defined scattering rings are present (Sect.~\ref{sec:DustScattering}). For instance, some intervening dust layers towards 
V404\,Cyg, which displayed a series of time variable rings in 2015, have been found to have steeper dust size distributions with respect to MRN. A population of large grains was also reported along this line of sight \citep[$ a> 0.15\,\mu$m,][]{Heinz2016,vp16,beardmore16}. An upper limit of $a<0.4$\,$\mu$m has been found along the diffuse sight lines of sources within 5\,kpc \citep{valencic15}. Very large grains (with an upper limit of $\sim$0.6\,$\mu$m) have been tentatively suggested for a line of sight near the Galactic center \citep[][]{jin17}. However, given the variety of grain size distributions available in the literature, fitting 1-D surface brightness profiles of dust scattering halos often yields inconclusive results due to a degeneracy between the grain size and dust cloud location when modeling the intensity profile \citep[][]{Costantini2005,xiang05,mao14,valencic15}. Using other markers of dust in the ISM, such as the optical extinction properties $A_V$ or $E(B-V)$, can assist in constraining the dust grain population models \citep{valencic15}. Comparing the dust scattering halo intensity profiles among different energy bands can also break this degeneracy \citep{CorralesPaerels2015}. \\

\noindent
{\bf Crystallinity}\\
As seen above (Fig.~\ref{f:size}, central panel), the sharpness of the XAFS near the edge may be an indicator, along side the edge shape and position, of absorption by dust in crystalline form. Recent modeling of the X-ray Si and Mg edges reports consistently the presence of crystalline dust, next to the amorphous component. Its contribution to the total dust was reported to be 4--35\% when only the Si\,K edge was considered \citep[][]{zeegers19}. A similar range of $\sim$7--27\% was found for a different set of sources when a simultaneous Mg and Si fitting was performed \citep[][]{rogantini20}. This is in apparent contradiction with studies in the IR \citep[e.g.][Sect.~\ref{par:star}]{kemper04,li07}. However X-rays and IR may have different diagnostic power, as the X-rays reach the short range domain (atomic distances), while the IR tests the long range interaction (molecular distances). More studies are necessary to corroborate this findings, however the presence of more crystalline structures than thought before may open new debate on crystalline dust formation and survival in the ISM. \\ 

\noindent
{\bf Abundance and depletion} \\
As seen in Sect.~\ref{par:attenuation}, the X-ray band does not host any distinctive hydrogen feature, but has to rely on the modeling of the soft energy cut off to determine the hydrogen column density. The evaluation of the metal abundances therefore depends on a reliable determination of the hydrogen along the line of sight and on the cosmic abundance chosen. In some circumstances \citep[e.g.][]{lee02,vanpeet09,grinberg15} additional cold material, possibly variable in time, may be associated to the immediate surrounding of the source, providing an apparent overabundance of metals along the line of sight.\\
The metal abundances are determined using the depth of the respective edges and the strength of their gas transitions. The absorption edge includes both gas and dust total contributions, providing an immediate measure of the amount of a given element. Adopting \citep[][]{lodders09} as reference for solar abundances, the values recently reported do not differ dramatically from the solar ones. A deviation of 10--20\% at most for iron and oxygen \citep[][]{pinto13,costantini12} and few percent for Mg and Si \citep[e.g.][]{rogantini20} has been found. The oxygen cold-gas content has been studied in detail in \citep{nicastro16,gatuzz16}. From the modeling of the \ion{O}{I} absorption line in a sample of Galactic X-ray sources spectra,  \citet{nicastro16} found the oxygen abundance  for the cold medium, residing mainly in the Galactic disk, to be slightly super-solar, using the oxygen abundance  prescribed in \citep{wilms00} ($A_{\rm O}^\odot=4.9\times10^{-4}$). 
Translating their result with the \citep{lodders09} oxygen abundance ($A_{\rm O}^\odot=6.0\times10^{-4}$), for an easier comparison, a value of $A_{\rm O}/A_{\rm O}^\odot\sim1.3$ is obtained. Using a similar approach, \citep{gatuzz16} report a wide range of abundance values, mostly sub-solar, along the line of sight of 24 X-ray binaries. It has to be noted that large uncertainties are associated to many of those measurements. Again, converting those values from the oxygen abundances used in that work \citep[$6.7\times10^{-4}$, ][]{gs98}, to the value in \citep{lodders09}, a range $A_{\rm O}/A_{\rm O}^\odot\sim0.25-1.5$, with a median of $\sim0.7$, is obtained.\\
The Fe abundance, derived from the Fe L edges, assuming gas as the only absorber, have been also studied in \citep{gatuzz16}. With large associated uncertainties, the range reported is $A_{\rm Fe}/A_{\rm Fe}^\odot\sim0.17-1.55$, with a median of $\sim0.65$. Note that the reference for the Fe solar abundance is very similar in this case for \citep{gs98} and \citep{lodders09}: $3.15\times10^{-5}$ and $3.25\times10^{-5}$, respectively. \\

Depletion values as measured in the X-rays generally confirm what hypothesized using longer wavelength techniques. In particular, when gas and dust components are used to model the Mg, Si, and Fe edges, it is systematically found that dust dominates the absorption. The Si inclusion in dust has been well constrained to be in the range 96--98\% for all high-column density lines of sight \citep{rogantini20}. For magnesium,  upper limits have been determined \citep[$>74-99$\%][]{rogantini20}. Iron has been also reported to be included almost totally in the dust phase \citep[e.g. $\sim$87\%][]{costantini12}. Oxygen has been studied along diffuse lines of sight and its depletion has been found to be moderate. Around 7--20\%  is the amount of oxygen in dust necessary to fit the O\,K edge, the rest being in gas form \citep[][]{pinto13,costantini12,psaradaki20}, consistently with what observed in optical/UV \citep[][]{jenkins09}. \\

\noindent
{\bf Dust spatial distribution} \\
Scattering halos offer a novel method to measure the spatial distribution of dust along the line of sight of a bright X-ray source. 
As described in Sect.~\ref{sec:DustScattering}, quiescent dust scattering halo profiles are smooth and offer limited insight on the exact position of intervening dust \citep[e.g.,][]{Smith2002}. This is because the two parameters, distance of the background source and the distance of the scattering dust cloud, are degenerate with the dust grain size distribution \citep[][]{predehlklose96}, with resulting uncertainties that may reach hundreds of parsecs \citep[e.g.,][]{mao14,xiang11}. 
Nonetheless, the latest survey of of 35 quiescent dust scattering halos imaged with the \textit{Chandra} and \textit{XMM-Newton} observatories finds that the majority of nearby ($D < 5$~kpc) ISM sight lines are well fit with single clouds \citep{Valencic2015}, while more distant sources required multiple clouds or size distributions. 
More precise measurements of the foreground spatial distribution of dust can be determined from dust scattering echoes, especially when combined with other tracers of interstellar dust such as CO maps \citep{Tiengo2010,Heinz2015} and stellar extinction studies \citep{Heinz2016}.
It is anticipated that dozens more high contrast dust scattering echoes could be imaged with the next generation of X-ray telescopes, for this type of study \citep{Corrales2019}.

\section{Future outlook}
\begin{figure}
\hspace{-0.7cm}
\hbox{
    \includegraphics[width=0.4\textwidth,angle=90]{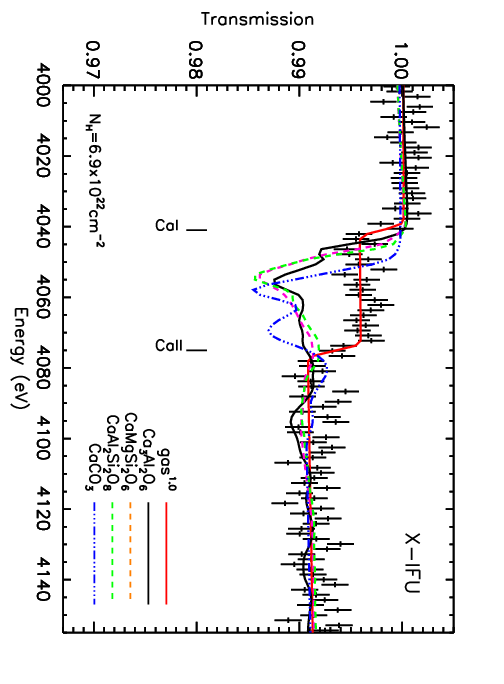}
     \hspace{-0.5cm}
     \includegraphics[width=0.4\textwidth,angle=90]{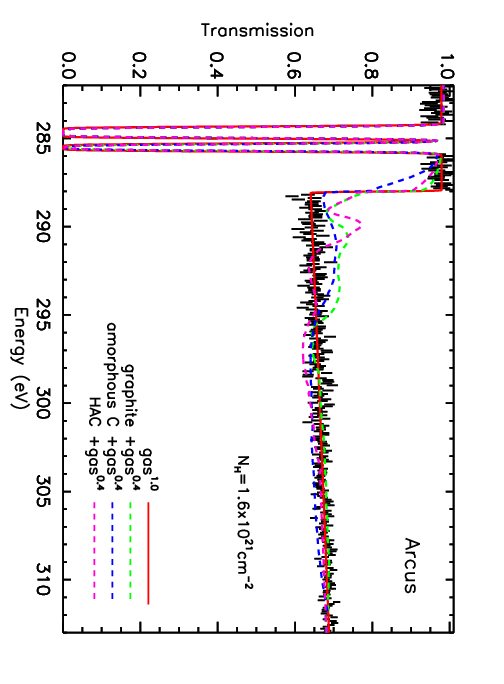}

 }
    \caption{\label{f:ca} Left panel: Athena-XIFU simulation of the Ca\,K edge for an intervening column density of $N_{\rm H}=6.9\times10^{22}$\,cm$^{-2}$. The simulated data (adopting only gas) are compared with models with different types of dust. While different types of silicates will be difficult to distinguish using the Ca K edge alone, other compounds will be easily disentangled. Adapted from \citet{costantini19}. Right panel: Arcus simulation of the carbon region, for a source with a moderate column density. The simulated data (adopting only gas) are compared with models with different types of dust and gas mixture. Carbon in gas is assumed to be 40\% of the total. The simulation shows that graphite and (hydrogenated) amorphous carbon can be disentangled. Adapted from \citet{costantini19}.}
\end{figure}

\begin{figure}
\centering 
     \includegraphics[width=0.8\textwidth,angle=0]{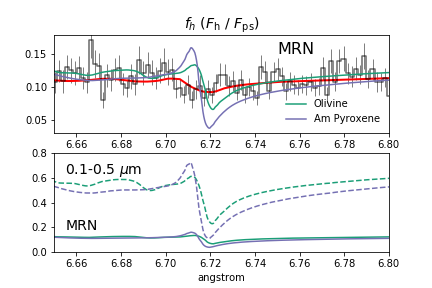}
     \caption{\label{fig:halo_ratio} Simulated ratio, using the XRISM-Resolve resolution, of scattering halo flux ($F_h$) to point source flux ($F_{\rm ps}$), accounting for the fraction of the scattering halo captured ($f_h$) by a small field-of-view. (Top) A zoom-in of the scattering features from silicate dust around the Si\,K edge, arising from an ISM column density of $N_{\rm H} \approx 4 \times 10^{22} {\rm cm}^{-2}$ and assuming an MRN \citep{Mathis1977} power-law distribution of dust grains with particle sizes less than $0.3~\mu {\rm m}$. The simulated data and red curve follow the theoretical cross-sections for silicate features from \citep{Draine2003}. The features modeled from laboratory data of \citet{Zeegers2017} are overlaid, for olivines (green) and amorphous pyroxenes (purple). (Bottom) Demonstration of the change in scattering halo flux from the contribution of large dust grains. The same ratio is plotted as in the top panel, for the MRN distribution of grains (solid curves). The same mass of dust, following a power-law slope of MRN but consisting of dust grains between 0.1 and 0.5~$\mu {\rm m}$ in radius, produces a much brighter scattering halo and XSFS features with a larger amplitude (dashed curves).}

\end{figure}

The X-ray observatories \xmm\ and \ch\ have enormously advanced the study of the ISM, thanks to the resolving power ($R=E/\Delta E\sim 400-1000$) of the grating instruments in the soft ($\sim$0.5--2\,keV) X-ray band. Future instruments, for example the calorimeters on board XRISM \citep{tashiro} and Athena \citep[][]{barret}, will explore at higher resolution the energy range with energies $> 1-2$\,keV. This will make accessible different absorption edges \citep[Al, S, Ca, Fe K edges][]{rogantini18,costantini19}, in addition to an even better view of the Mg and Si\,K edges. 
The energy resolution, $\Delta E$, will be 5 and 2.5\,eV for XRISM-Resolve and Athena-XIFU, respectively. This, combined with a high effective area in the 2--10\,keV range, will allow the study of XAFS even for the shallower edges expected for e.g. Ca and Al (Fig.~\ref{f:ca}). The new energy window will allow the X-ray exploration of very-high column density molecular environments, near the Galactic center. In particular, the iron inclusion in those environments will be studied, by means of the Fe\,K edge at 7.1\,keV \citep{rogantini18}. Absorption edges from Mg and Si will be explored into greater detail, allowing a deeper study on grain size distribution \citep{xrism_wp}.\\ 
A higher resolving power, as proposed for the concept mission Arcus \citep[$R\sim3800$,][]{smith20}, in the soft energy band would allow to study in detail, alongside the iron L edges,  
the oxygen K edge, rich in absorption features of gas and dust \citep[e.g.][]{juett04, costantini12, psaradaki20} and the carbon edge at 0.28\,keV (Fig.~\ref{f:ca}, right panel). The study of this edge, with high resolution spectroscopy, would reveal the physical characteristics of one of the major components of ISM, as graphite would show distinctive features with respect to amorphous carbon or HAC \citep[][and Sect.~\ref{par:star}]{costantini19}.\\  
The high sensitivity provided by both future calorimeters and CCDs-imaging instruments \citep[][]{tashiro,nandra} will allow to study many more, fainter scattering halos. Five Galactic X-ray sources have produced the brightest dust ring echoes to date \citep{Tiengo2010, Heinz2015, Heinz2016, Kalemci2018, Nobukawa2020}. The next generation of X-ray observatories, with 50-80 times the sensitivity of current instruments, are expected to capture high contrast dust ring echoes at about $30$ times the frequency of current X-ray observatories \citep{Corrales2019}. \\
The bright sources will offer to routinely perform spatially resolved halo spectroscopy, that reveals the chemical properties of dust alone \citep{decourchelle13,xrism_wp}. In Fig.~\ref{fig:halo_ratio}, an example of how spatially resolved scattering halo spectra will be observed is shown. Once the contribution of the central source spectrum has been divided out, the scattering halo spectrum will reveal the XSFS features (Sect.~\ref{par:xafs}). Those will allow us to determine both the chemistry  (Fig.~\ref{fig:halo_ratio}, upper panel) and the dust size distribution (lower panel) of virtually any line of sight displaying a scattering halo.\\

\noindent
{\it Acknowledgements}\\
The authors wish to thank S.\,Zeegers, D.\,Rogantini, I.\,Psaradaki, R.\,Waters and S.\,Heinz for interesting discussions, help with the figures, and for their insightful comments on this manuscript.

\end{document}